\documentstyle[11pt,aaspp4,tighten]{article}
\eqsecnum

\def\etal{{\it et al.~}}
\def\eg{e.g.,~}

\begin{document}

\title{A Multi-dimensional Code for Isothermal Magnetohydrodynamic Flows
       in Astrophysics\altaffilmark{5}}
\author{Jongsoo Kim\altaffilmark{1}, Dongsu Ryu\altaffilmark{2},
    T. W. Jones\altaffilmark{3}, and Seung Soo Hong\altaffilmark{4}}

\altaffiltext{1}
{Korea Astronomy Observatory, San 36-1, Hwaam-Dong, Yusong-Ku,
Taejon 305-348, Korea and Department of Astronomy, Seoul National
University, Seoul 151-742, Korea: jskim@hanul.issa.re.kr}
\altaffiltext{2}
{Department of Astronomy \& Space Science, Chungnam National University,
Taejon 305-764, Korea:\\ryu@canopus.chungnam.ac.kr}
\altaffiltext{3}
{Department of Astronomy, University of Minnesota, Minneapolis, MN 55455:
twj@mail.msi.umn.edu} 
\altaffiltext{4}
{Department of Astronomy, Seoul National University, Seoul 151-742, Korea:
sshong@astroism.snu.ac.kr}
\altaffiltext{5}
{To appear in the Astrophysical Journal}

\begin{abstract}

We present a multi-dimensional numerical code to solve isothermal
magnetohydrodynamic (IMHD) equations for use in modeling astrophysical
flows.
First, we have built a one-dimensional code which is based on an
explicit finite-difference method on an Eulerian grid, called the
total variation diminishing (TVD) scheme.
The TVD scheme is a second-order-accurate extension of the Roe-type
upwind scheme.
Recipes for building the one-dimensional IMHD code, including the
normalized right
and left eigenvectors of the IMHD Jacobian matrix, are presented.
Then, we have extended the one-dimensional code to a multi-dimensional
IMHD code through a Strang-type dimensional splitting.
In the multi-dimensional code, an explicit cleaning step has been
included to eliminate non-zero $\nabla\cdot B$ at every time step.

To test the code, IMHD shock tube problems, which encompass all the
physical IMHD structures, have been constructed.
One-dimensional and two-dimensional shock tube tests have
shown that the code captures
all the structures correctly without producing noticeable oscillations.
Strong shocks are resolved sharply, but weaker shocks spread more.
Numerical dissipation (viscosity and resistivity) has been estimated
through the decay test of a two-dimensional Alfv\'{e}n wave.
It has been found to be slightly smaller than that of the adiabatic
magnetohydrodynamic code based on the same scheme.
As an example of astrophysical applications, we have simulated the
nonlinear evolution of the two-dimensional Parker instability under
a uniform gravity.

\end{abstract}


\section{INTRODUCTION} 

Over the last two decades, conservative upwind differencing schemes have
proven to be very efficient for solving {\it adiabatic} hydrodynamic and
magnetohydrodynamic (MHD) equations.  These methods generally depend on
the calculated estimates of mass, momentum and energy fluxes as well as
magnetic field flux across cell boundaries based on the so-called
``Riemann'' solutions from the basic conservation laws.
Examples for hydrodynamics include extensions of Godunov's scheme
(Godunov 1959), such as the MUSCL scheme (Van Leer 1979) and the PPM
scheme (Colella \& Woodward 1984), as well as those based on approximate
flow eigenstates such as the Roe's scheme (Roe 1981), the TVD scheme
(Harten 1983) and the ENO scheme (Harten \etal 1987).
Works for MHD include Brio \& Wu (1988), Zachary \& Colella (1992),
Zachary, Malagoli, \& Colella (1994), Dai \& Woodward (1994a,1994b),
Ryu \& Jones (1995) (RJ, hereafter), Ryu, Jones, \& Frank (1995) (RJF,
hereafter), Powell \etal (1995), and Roe \& Balsara (1996).
Brio \& Wu applied the Roe's approach to the MHD equations.
Zachary and collaborators used the BCT scheme to estimate fluxes for the
MHD conservation equations.
Dai \& Woodward adapted the PPM scheme to MHD.
Ryu and collaborators extended the Harten's TVD scheme to MHD.
Powell and collaborators developed a Roe-type Riemann solver with
an {\it eight}-wave structure for MHD.
Roe \& Balsara constructed one variety of linearized Riemann solutions
for MHD.
The upwind schemes generally share an ability to sharply and cleanly
define fluid discontinuities, especially shocks, and exhibit a robustness
that makes them broadly applicable.

The assumption of {\it adiabatic} flows holds in the limit where
cooling is negligible or the cooling time scale is much larger than the 
dynamical time scale.  However, in the other limit where the cooling time 
scale is much shorter than the dynamical time scale, the assumption of 
{\it isothermal} flows becomes physically more valid
(\eg Draine \& McKee 1993 and references therein).
Of course, if cooling time scale is comparable to dynamical time scale,
cooling should be considered explicitly.

Usually, numerical simulations of isothermal flows are made with
adiabatic codes by setting the adiabatic index, $\gamma$, close to unity.
Truelove \etal (1998) showed that with $\gamma$ as close as to unity
as $1.001$, their adiabatic hydrodynamic code can follow isothermal collapse
without any significant deterioration of accuracy.
We also observed that with $\gamma=1.001$ the adiabatic TVD MHD code (RJ)
captures structures in isothermal magnetohydrodynamic (IMHD) shock tubes
without noticeable error.
Yet, it is desirable to build codes specifically for isothermal flows,
since those codes are {\it simpler} and {\it faster} than adiabatic ones.
That is because the energy conservation equation need not to be solved in
isothermal codes.
As a result, the entropy mode, which carries the contact discontinuity, need
not to be considered.
In the current paper we describe an IMHD code based on Harten's TVD
scheme.  It is the same scheme that was used for the adiabatic MHD code
in RJ and RJF.
Balsara (1998b) developed an IMHD code also based on an upwind scheme,
but his scheme is different from ours.

In \S 2, we give recipes for the development of one- and multi-dimensional 
IMHD codes. In \S 3, we present the results of tests
that include one-dimensional and two-dimensional shock tube problems, the
decay of an Alfv\'{e}n wave,
and the nonlinear evolution of the Parker instability under a
uniform gravity.  Conclusions follow in \S 4.

\section{NUMERICAL SCHEME}

\subsection{The Equations for Isothermal Magnetohydrodynamics}

MHDs describes the behavior of the combined system of a conducting fluid
and magnetic fields in the limit that the displacement current and
the separation between ions and electrons are neglected.  So the MHD
equations represent coupling of the equations of
fluid dynamics with Maxwell's equations of electrodynamics.
By ignoring the effects
of electrical resistivity, viscosity, and thermal conductivity,
and imposing isothermality on the conducting fluid, we get
the following IMHD equations:
\begin{equation}
\frac{\partial \rho}{\partial t} + \nabla \cdot (\rho v) = 0,
\end{equation}
\begin{equation}\label{momentum}
\frac{\partial v}{\partial t} 
+ v \cdot \nabla v 
+ \frac{1}{\rho}\nabla(a^2\rho)
-\frac{1}{\rho}(\nabla \times B)\times B =0,
\end{equation}
\begin{equation}
\frac{\partial B}{\partial t} - 
\nabla \times (v \times B) = 0,
\end{equation}
with an additional constraint 
\begin{equation}\label{divb}
\nabla \cdot B=0,
\end{equation}
for the absence of magnetic monopoles.  
Here $a$ is an isothermal sound speed, and other notations have
their usual meanings.  We incorporate a factor of $1/\sqrt{4\pi}$
into the definition of $B$ so that the factor of $4\pi$ does not appear 
in Eq.~(\ref{momentum}).  

In Cartesian coordinates, the above equations are written in 
a conservative form as
\begin{equation}\label{3dimhd}
\frac{\partial q}{\partial t} 
+ \frac{\partial F_x}{\partial x}
+ \frac{\partial F_y}{\partial y}
+ \frac{\partial F_z}{\partial z} = 0,
\end{equation}
\begin{equation}\label{qfx}
q = \left( \begin{array}{c}
             \rho     \\
             \rho v_x \\
             \rho v_y \\
             \rho v_z \\
             B_x      \\
             B_y      \\      
             B_z      
           \end{array}
    \right), \;
F_x = \left( \begin{array}{c}
             \rho v_x                                       \\
             \rho v_x^2 + a^2 \rho + (B_y^2+B_z^2-B_x^2)/2  \\
             \rho v_x v_y - B_x B_y                         \\
             \rho v_x v_z - B_x B_z                         \\
             0                                              \\
             v_x B_y - v_y B_x                              \\
             v_x B_z - v_z B_x
           \end{array}
    \right),
\end{equation} 
with $F_y$ and $F_z$ obtained by properly permuting indices.
With the state vector $q$ and the flux functions $F_x(q)$, 
$F_y(q)$, and $F_z(q)$, the Jacobian matrices, 
$A_x(q) = \partial F_x/\partial q$, 
$A_y(q) = \partial F_y/\partial q$, and  
$A_z(q) = \partial F_z/\partial q$ are formed.  A system is called
{\it hyperbolic} if all the eigenvalues of the Jacobian matrices
are real and distinct and the corresponding set of right eigenvectors is
complete (Jeffrey \& Taniuti 1964).  The system of the ideal, adiabatic
MHD equations is known as {\it non-strictly hyperbolic}, since some
eigenvalues coincide at some points (Brio \& Wu 1988; Roe \& Balsara 1996). 
The eigen-structure of the IMHD equations, which is presented in the next
subsection, is very similar to that of the adiabatic ones.    
It is easy to show that the IMHD equations also form a non-strictly
hyperbolic system.

\subsection{One-Dimensional Code}

Our strategy for developing a one-dimensional IMHD code is based
the TVD scheme (Harten 1983) which was devised to improve 
the first-order-accurate Roe's upwind scheme (Roe 1981) into a
second-order-accurate one.
For it, we derive the eigenvalues and eigenvectors of the system 
of the IMHD equations, which are given below.  
With the eigenvalues and eigenvectors, it is straightforward to apply  
the construction procedure for the one-dimensional adiabatic MHD code 
(RJ) to an isothermal analogue.
Even though the procedure is described in RJ,
we here repeat it to make this paper self-contained.    
Special attention is given to the orthonormal eigenvectors of the 
system of the IMHD equations.

We consider, as an example, plane-symmetric, one-dimensional flows
exhibiting variation along the $x$-direction.  
Then $y$- and $z$-derivatives in Eq.~(\ref{3dimhd})
are zero, and we have the one-dimensional IMHD equations
\begin{equation}\label{1dimhd}
\frac{\partial q}{\partial t} 
+ \frac{\partial F_x}{\partial x} = 0,
\end{equation}
where $q$ and $F_x$ are defined in Eq.~(\ref{qfx}).  
The fifth equation in the system of Eqs.~(\ref{1dimhd}) is 
$\frac{\partial}{\partial t} B_x = 0$, and the constraint in
Eq.~(\ref{divb}) is $\frac{\partial}{\partial x} B_x =0$. 
These imply that initially $B_x$ should
be spatially constant and be kept constant during the evolution
of the flow.  So we need not include the equation for $B_x$ in
a one-dimensional code.

The Jacobian matrix $\partial F_x/\partial q$ of the system of 
Eqs.~(\ref{1dimhd}) is given by
\begin{equation}
A_x = \left( \begin{array}{cccccc}
              0 & 1 & 0 & 0 & 0 & 0                                      \\
              a^2-v_x^2 & 2v_x & 0 & 0 & b_y\sqrt{\rho} & b_z\sqrt{\rho} \\
              -v_xv_y   & v_y  & v_x & 0   & -b_x\sqrt{\rho} & 0         \\
              -v_xv_z   & v_z  &  0  & v_x & 0 & -b_x\sqrt{\rho}         \\
              -v_x\frac{b_y}{\sqrt{\rho}}+v_y\frac{b_x}{\sqrt{\rho}} & 
               \frac{b_y}{\sqrt{\rho}} & -\frac{b_x}{\sqrt{\rho}} & 
               0 & v_x & 0                                            \\   
              -v_x\frac{b_z}{\sqrt{\rho}}+v_z\frac{b_x}{\sqrt{\rho}} & 
               \frac{b_z}{\sqrt{\rho}} & 0 & 
               -\frac{b_x}{\sqrt{\rho}} & 0 & v_x
           \end{array}
     \right),
\end{equation}
where $b_{x,y,z} = B_{x,y,z}/\sqrt{\rho}$.
The six eigenvalues in non-decreasing order are
\begin{equation}
a_1 = v_x - c_f,
\end{equation}
\begin{equation}
a_2 = v_x - c_a,
\end{equation}
\begin{equation}
a_3 = v_x - c_s,
\end{equation}
\begin{equation}
a_4 = v_x + c_s,
\end{equation}
\begin{equation}
a_5 = v_x + c_a,
\end{equation}
\begin{equation}
a_6 = v_x + c_f,
\end{equation}
where $c_f$, $c_a$, $c_s$ are the fast, Alfv\'{e}n, and slow characteristic
speeds, respectively.  There is no entropy mode for the IMHD equations.  
The quantities $a_1, \cdots, a_6$ represent the six speeds
with which information is propagated locally by three MHD wave families.
The three characteristic speeds are expressed as
\begin{equation}
c_f = \left\{\frac{1}{2}
      \left[a^2+b_x^2+b_y^2+b_z^2+\sqrt{(a^2+b_x^2+b_y^2+b_z^2)^2-4a^2b_x^2}
      \right]\right\}^{1/2},
\end{equation}
\begin{equation}
c_a = |b_x|,
\end{equation}
\begin{equation}
c_s = \left\{\frac{1}{2}
      \left[a^2+b_x^2+b_y^2+b_z^2-\sqrt{(a^2+b_x^2+b_y^2+b_z^2)^2-4a^2b_x^2}
      \right]\right\}^{1/2}.
\end{equation}
The six right eigenvectors corresponding to the six eigenvalues are
\begin{equation}
R_{v_x \pm c_f} = \left( \begin{array}{c}
                           1                                          \\
                           v_x \pm c_f                                \\
                           v_y \mp \frac{c_f b_x b_y}{c_f^2-b_x^2}    \\
                           v_z \mp \frac{c_f b_x b_z}{c_f^2-b_x^2}    \\
                           \frac{c_f^2 b_y}{(c_f^2-b_x^2)\sqrt{\rho}} \\
                           \frac{c_f^2 b_z}{(c_f^2-b_x^2)\sqrt{\rho}}
                         \end{array} 
                  \right),
\end{equation}
\begin{equation}
R_{v_x \pm c_a} = \left( \begin{array}{c}
                           0                       \\
                           0                       \\
                           \mp b_z {\rm sign}(b_x) \\
                           \pm b_y {\rm sign}(b_x) \\
                           \frac{b_z}{\sqrt{\rho}} \\
                          -\frac{b_y}{\sqrt{\rho}} \\
                         \end{array} 
                  \right),
\end{equation}
\begin{equation}
R_{v_x \pm c_s} = \left( \begin{array}{c}
                           1                                          \\
                           v_x \pm c_s                                \\
                           v_y \mp \frac{c_s b_x b_y}{c_s^2-b_x^2}    \\
                           v_z \mp \frac{c_s b_x b_z}{c_s^2-b_x^2}    \\
                           \frac{c_s^2 b_y}{(c_s^2-b_x^2)\sqrt{\rho}} \\
                           \frac{c_s^2 b_z}{(c_s^2-b_x^2)\sqrt{\rho}}
                        \end{array}
                  \right). 
\end{equation}
Near the point where either $b_x=0$ or $b_y=b_z=0$, the above
right eigenvectors are not well defined, with some elements becoming singular.
By re-normalizing the eigenvectors, the singularities can be removed.  
The renormalized right eigenvectors are
\begin{equation}
R_{v_x \pm c_f} = \left( \begin{array}{c}
                           \alpha_f                                 \\
                           \alpha_f(v_x \pm c_f)                    \\
                           \alpha_f v_y \mp \alpha_s \beta_y b_x    \\
                           \alpha_f v_z \mp \alpha_s \beta_z b_x    \\
                           \frac{\alpha_s \beta_y c_f}{\sqrt{\rho}} \\
                           \frac{\alpha_s \beta_z c_f}{\sqrt{\rho}} 
                         \end{array} 
                  \right),
\end{equation}
\begin{equation}
R_{v_x \pm c_a} = \left( \begin{array}{c}
                           0                           \\
                           0                           \\
                           \mp \beta_z {\rm sign}(b_x) \\
                           \pm \beta_y {\rm sign}(b_x) \\
                           \frac{\beta_z}{\sqrt{\rho}} \\
                          -\frac{\beta_y}{\sqrt{\rho}}
                         \end{array} 
                  \right),
\end{equation}
\begin{equation}
R_{v_x \pm c_s} = \left( \begin{array}{c}
                           \alpha_s                                          \\
                           \alpha_s(v_x \pm c_s)                             \\
                           \alpha_s v_y \pm \alpha_f\beta_y a {\rm sign}(b_x)\\
                           \alpha_s v_z \pm \alpha_f\beta_z a {\rm sign}(b_x)\\
                          -\frac{\alpha_f\beta_y a^2}{c_f\sqrt{\rho}}        \\
                          -\frac{\alpha_f\beta_z a^2}{c_f\sqrt{\rho}}
                         \end{array} 
                  \right),
\end{equation}
where $\alpha$'s and $\beta$'s are defined by 
\begin{equation}
\alpha_f = \frac{\sqrt{c_f^2-b_x^2}}{\sqrt{c_f^2-c_s^2}},
\end{equation}
\begin{equation}
\alpha_s = \frac{\sqrt{c_f^2-a^2}}{\sqrt{c_f^2-c_s^2}},
\end{equation}
\begin{equation}
\beta_y = \frac{b_y}{\sqrt{b_y^2+b_z^2}},
\end{equation}
\begin{equation}
\beta_z = \frac{b_z}{\sqrt{b_y^2+b_z^2}}.
\end{equation}
At the points where $b_y=b_z=0$, $\beta$'s are defined as
a limiting value, i.e.,
\begin{equation}
\beta_y = \beta_z = \frac{1}{\sqrt{2}}.
\end{equation}
Similarly, at the point where $b_y=b_z=0$ and $b_x^2=a^2$, $\alpha$'s are
defined as
\begin{equation}
\alpha_f=\alpha_s=1.
\end{equation}

The left eigenvectors, which are orthonormal to the right eigenvectors,
$L_{l} \cdot R_{m} = \delta_{lm}$, are
\begin{equation}
L_{v_x \pm c_f} = (l_{v_x \pm c_f}^{(1)}, l_{v_x \pm c_f}^{(2)}, 
                   l_{v_x \pm c_f}^{(3)}, l_{v_x \pm c_f}^{(4)}, 
                   l_{v_x \pm c_f}^{(5)}, l_{v_x \pm c_f}^{(6)}),
\end{equation}
\begin{equation}
l_{v_x \pm c_f}^{(1)} = 
         \frac{1}{\theta_1} \alpha_f a^2
     \pm \frac{1}{\theta_2}
         \left[ -\alpha_f a v_x 
                +\alpha_s c_s (\beta_yv_y+\beta_zv_z){\rm sign}(b_x) 
         \right],
\end{equation}
\begin{equation}
l_{v_x \pm c_f}^{(2)} = 
     \pm \frac{1}{\theta_2} \alpha_f a,
\end{equation}
\begin{equation}
l_{v_x \pm c_f}^{(3)} = 
     \mp \frac{1}{\theta_2} 
         \alpha_s \beta_y c_s {\rm sign}(b_x), 
\end{equation}
\begin{equation}
l_{v_x \pm c_f}^{(4)} = 
     \mp \frac{1}{\theta_2} 
         \alpha_s \beta_z c_s {\rm sign}(b_x), 
\end{equation}
\begin{equation}
l_{v_x \pm c_f}^{(5)} = 
         \frac{1}{\theta_1} \alpha_s \beta_y c_f \sqrt{\rho},
\end{equation}
\begin{equation}
l_{v_x \pm c_f}^{(6)} = 
         \frac{1}{\theta_1} \alpha_s \beta_z c_f \sqrt{\rho},
\end{equation}
\begin{equation}
L_{v_x \pm c_a} = (l_{v_x \pm c_a}^{(1)}, l_{v_x \pm c_a}^{(2)}, 
                   l_{v_x \pm c_a}^{(3)}, l_{v_x \pm c_a}^{(4)}, 
                   l_{v_x \pm c_a}^{(5)}, l_{v_x \pm c_a}^{(6)}), 
\end{equation}
\begin{equation}
l_{v_x \pm c_a}^{(1)} = 
     \pm \frac{1}{2}(\beta_zv_y-\beta_yv_z){\rm sign}(b_x),
\end{equation}
\begin{equation}
l_{v_x \pm c_a}^{(2)} = 0, 
\end{equation}
\begin{equation}
l_{v_x \pm c_a}^{(3)} = \mp \frac{1}{2}\beta_z{\rm sign}(b_x), 
\end{equation}
\begin{equation}
l_{v_x \pm c_a}^{(4)} = \pm \frac{1}{2}\beta_y{\rm sign}(b_x), 
\end{equation}
\begin{equation}
l_{v_x \pm c_a}^{(5)} =  \frac{1}{2}\beta_z\sqrt{\rho},
\end{equation}
\begin{equation}
l_{v_x \pm c_a}^{(6)} = -\frac{1}{2}\beta_y\sqrt{\rho},
\end{equation}
\begin{equation}
L_{v_x \pm c_s} = (l_{v_x \pm c_s}^{(1)}, l_{v_x \pm c_s}^{(2)}, 
                   l_{v_x \pm c_s}^{(3)}, l_{v_x \pm c_s}^{(4)}, 
                   l_{v_x \pm c_s}^{(5)}, l_{v_x \pm c_s}^{(6)}),
\end{equation}
\begin{equation}
l_{v_x \pm c_s}^{(1)} = 
         \frac{1}{\theta_1} \alpha_s c_f^2
     \mp \frac{1}{\theta_2}
         \left[ \alpha_s c_a v_x 
               +\alpha_f c_f (\beta_yv_y+\beta_zv_z){\rm sign}(b_x) 
         \right],
\end{equation}
\begin{equation}
l_{v_x \pm c_s}^{(2)} = 
     \pm \frac{1}{\theta_2} \alpha_s c_a,
\end{equation}
\begin{equation}
l_{v_x \pm c_s}^{(3)} = 
     \pm \frac{1}{\theta_2} 
         \alpha_f \beta_y c_f {\rm sign}(b_x),
\end{equation}
\begin{equation}
l_{v_x \pm c_s}^{(4)} = 
     \pm \frac{1}{\theta_2} 
         \alpha_f \beta_z c_f {\rm sign}(b_x), 
\end{equation}
\begin{equation}
l_{v_x \pm c_s}^{(5)} = 
       - \frac{1}{\theta_1} \alpha_f \beta_y c_f \sqrt{\rho},
\end{equation}
\begin{equation}
l_{v_x \pm c_s}^{(6)} = 
       - \frac{1}{\theta_1} \alpha_f \beta_z c_f \sqrt{\rho},
\end{equation}
where
\begin{equation}
\theta_1 =  2(\alpha_f^2a^2 + \alpha_s^2c_f^2),
\end{equation}
\begin{equation}
\theta_2 =  2(\alpha_f^2c_fa +\alpha_s^2c_ac_s).
\end{equation}

Some elements in the normalized right and left eigenvectors are not 
continuous.  In order to force them to be continuous,
\begin{equation}
{\rm sign}(b_T) = \left\{
\begin{array}{ccccccc}
      1, & \mbox{if} & b_y>0 & \mbox{or} & b_y=0 & \mbox{and} & b_z>0 \\
     -1, & \mbox{if} & b_y<0 & \mbox{or} & b_y=0 & \mbox{and} & b_z<0,
\end{array}
                  \right.
\end{equation}
is multiplied to $R_{v_x \pm c_s}$ and $L_{v_x \pm c_s}$
if $a^2>c_a^2$, and to $R_{v_x \pm c_f}$ and $L_{v_x \pm c_f}$
if $a^2<c_a^2$.

Note that our eigenvectors have a different form from those derived in
Balsara (1998a).  The exact form of the eigenvectors does not matter,
once all the singular points are taken care of.

Here, we use the conventional indices.  The superscript $n$ represents
the time step.  The subscript $i$ indicates quantities at the cell
center, while $i+\case{1}{2}$ marks those at the right-hand cell
boundary.  The subscript $k$ represents the characteristic fields,
with the order that $k=1$ is for the field associated with eigenvalue
$v_x-c_f$, $k=2$ for the field with $v_x-c_a$, $k=3$ for the field
with $v_x-c_s$, $k=4$ for the field with $v_x+c_s$, $k=5$ for the
field with $v_x+c_a$, and finally $k=6$ for the field with $v_x+c_f$. 

An important step in the Roe's scheme (1981) is to determine a Roe matrix
$\bar{A}_{x,i+1/2}(q_{i},q_{i+1})$ at the cell boundary from the adjacent 
state vectors, which satisfies the Roe's suggested properties.  
One of them is $F_{x,i+1}-F_{x,i} = \bar{A}_{x,i+1/2}(q_{i+1}-q_{i})$.  
For the systems of the adiabatic and isothermal hydrodynamic equations, 
there exists a Roe matrix evaluated at the 
$\sqrt{\rho}$-weighted average state (Roe 1981; LeVeque 1997).  
For the system of the adiabatic MHD
equations, there is, however, no simple form of the Roe matrix except 
for the case with an adiabatic index $\gamma=2$ (Brio \& Wu 1988).  
We have failed to find a simple form of the Roe matrix for the system of the 
IMHD equations, too.  So we use an arithmetic averaging for the flow
quantities at the cell boundary in the IMHD code,
which was shown to work well in the adiabatic MHD code (RJ):
\begin{equation}
\rho_{i+1/2} = \frac{\rho_{i}+\rho_{i+1}}{2},
\end{equation}
\begin{equation}
v_{x,i+1/2} = \frac{v_{x,i}+v_{x,i+1}}{2},
\end{equation}
\begin{equation}
v_{y,i+1/2} = \frac{v_{y,i}+v_{y,i+1}}{2},
\end{equation}
\begin{equation}
v_{z,i+1/2} = \frac{v_{z,i}+v_{z,i+1}}{2},
\end{equation}
\begin{equation}
B_{y,i+1/2} = \frac{B_{y,i}+B_{y,i+1}}{2},
\end{equation}
\begin{equation}
B_{z,i+1/2} = \frac{B_{z,i}+B_{z,i+1}}{2}.
\end{equation}

The state vector $q$ at the cell center is updated by calculating the 
modified fluxes $\bar{f}_{x}$ at the cell boundaries as follows:
\begin{equation}
L_x q_{i}^{n+1} = q_{i}^{n} 
- \frac{\Delta t^n}{\Delta x}(\bar{f}_{x,i-1/2}-\bar{f}_{x,i+1/2}),
\end{equation}
\begin{equation}
\bar{f}_{x,i+1/2}=\frac{1}{2}[F_{x}(q_{i}^{n})+F_{x}(q_{i+1}^{n})]
-\frac{\Delta x}{2\Delta t^{n}} 
 \sum_{k=1}^{6} \beta_{k,i+1/2} R_{k,i+1/2}^{n},
\end{equation}
\begin{equation}
\beta_{k,i+1/2}= Q_{k}
\left(\frac{\Delta t^n}{\Delta x} a_{k,i+1/2}^{n}+\gamma_{k,i+1/2}\right)
\alpha_{k,i+1/2} - (g_{k,i}+g_{k,i+1}),
\end{equation}
\begin{equation}
\alpha_{k,i+1/2} = L_{k,i+1/2}^{n} \cdot (q_{i+1}^n-q_{i}^{n}),
\end{equation}
\begin{equation}
\gamma_{k,i+1/2} = \left\{
\begin{array}{ccc}
       \frac{g_{k,i+1}-g_{k,i}}{\alpha_{k,i+1/2}} & \mbox{for} & 
       \alpha_{k,i+1/2} \neq 0,  \\
       0 & \mbox{for} & \alpha_{k,i+1/2} = 0,
\end{array}
                   \right.
\end{equation}
\begin{equation}
g_{k,i}=\mbox{sign}(\tilde{g}_{k,i+1/2})
\mbox{max}[0,\mbox{min}\{|\tilde{g}_{k,i+1/2}|,
                         \tilde{g}_{k,i-1/2}\mbox{sign}(\tilde{g}_{k,i+1/2})\}]
\end{equation}
\begin{equation}
\tilde{g}_{k,i+1/2}=\frac{1}{2}
\left[Q_k\left(\frac{\Delta t^{n}}{\Delta x}a_{k,i+1/2}^{n}\right)
      -\left(\frac{\Delta t^{n}}{\Delta x}a_{k,i+1/2}^{n}\right)^{2}\right]
\alpha_{k,i+1/2},
\end{equation}
\begin{equation}
Q_{k}(\chi) = \left\{
\begin{array}{ccc}
      \frac{\chi^2}{4\epsilon_{k}} + \epsilon_{k} & \mbox{for} &
      |\chi| < 2\epsilon_{k}, \\
      |\chi| & \mbox{for} & |\chi| \geq 2\epsilon_{k}.
\end{array}
              \right.
\end{equation}
Since the use of {\it contact steepener} and
{\it rotational steepener} produces spurious numerical oscillations
in the adiabatic MHD code (RJ; RJF), we do not include
the rotational steepener in the IMHD code.   
The time step size $\Delta t^{n}$ is restricted by the usual Courant condition
for stability, $\Delta t^{n} = C_{\rm cour}\Delta x / {\rm Max} 
(|v_{x,i+1/2}^{n}|+c_{f,i+1/2}^{n})$ with $C_{\rm cour} < 1$.

\subsection{Multi-dimensional Code}

We extend the one-dimensional IMHD code to more dimensions by using
a Strang-type directional splitting (Strang 1968). Here we
explain, as an example, the implementation of it in 
two-dimensional plane-parallel geometry.
The two-dimensional IMHD equations written in the conservative
form (Eq.~[\ref{3dimhd}]) can be split into 
\begin{eqnarray}
L_x~(x\mbox{-sweep}): & q_t + \frac{\partial F_x}{\partial x} = 0,  \\
L_y~(y\mbox{-sweep}): & q_t + \frac{\partial F_y}{\partial y} = 0.
\end{eqnarray}
In a time step, we update the state vector $q(x,y)$
along the $x$-direction with
$y$ fixed, followed along the $y$-direction with $x$ fixed,
\begin{equation}
q^{n+1} = L_y L_x q^{n}.
\end{equation}
In order to maintain a second-order accuracy in time, the order of directional
sweeps is permuted in the next time step by  $L_x L_y$.
The time step size, $\Delta t$, is calculated at the start of the one complete
sequence of $L_x L_y$ $L_y L_x$ and fixed through the sequence.

In multi-dimensional simulations, numerical solutions may not satisfy
$\nabla \cdot B=0$ due to discretization errors.
Brackbill \& Barnes (1980) pointed out that errors of non-zero
$\nabla \cdot B$ appear as a force parallel to the field.
Non-zero $\nabla \cdot B$ can be removed, for instances, either by
incorporating an explicit {\it divergence cleaning} method as described in RJF
or by implementing a scheme similar to the {\it constrained transport}
scheme (Evans \& Hawley 1988) which was described in details for the
adiabatic MHD code in Ryu \etal (1998).
Tests in the next section have been done using the explicit divergence
cleaning method, and the next two paragraphs describe it briefly.

At the beginning of MHD simulations, $\nabla \cdot B = 0$ is satisfied.
The updated magnetic field $B$, which is not in general divergence-free,
can be decomposed as into two parts,
\begin{equation}\label{HP}
B = - \nabla \phi + \nabla \times V,  
\end{equation}
where $\phi$ and $V$ are scalar and vector functions respectively. 
Then the corrected magnetic field defined as $B^c = B + \nabla \phi$
becomes divergence-free.  So the problem of the divergence-cleaning is 
reduced to find $\phi$, which is described by the Poisson equation
\begin{equation}\label{Poisson}
\nabla^2 \phi = - \nabla \cdot B.
\end{equation}
In two-dimensional Cartesian geometry, for instance, the following finite
difference representations
\begin{equation}
B_{x,i,j}^c = B_{x,i,j} + \frac{\phi_{i+1,j}-\phi_{i-1,j}}{2\Delta x},
\end{equation}
\begin{equation}
B_{y,i,j}^c = B_{y,i,j} + \frac{\phi_{i,j+1}-\phi_{i,j-1}}{2\Delta y},
\end{equation}
together with
\begin{equation}\label{FD}
\frac{\phi_{i+2,j}-2\phi_{i,j}+\phi_{i-2,j}}{(2\Delta x)^2} +
\frac{\phi_{i,j+2}-2\phi_{i,j}+\phi_{i,j-2}}{(2\Delta y)^2} =
- \left( \frac{B_{x,i+1,j}-B_{x,i-1,j}}{2\Delta x} 
      +\frac{B_{y,i,j+1}-B_{y,i,j-1}}{2\Delta y} \right),
\end{equation}
ensures
\begin{equation}
\frac{B_{x,i+1,j}^c-B_{x,i-1,j}^c}{2\Delta x} +
\frac{B_{y,i,j+1}^c-B_{x,i,j-1}^c}{2\Delta y} = 0,
\end{equation}
within machine round-off error.
Extensions to the three-dimension and/or to other geometry are
straightforward.

Eq.~(\ref{FD}) is solved with boundary conditions specific to problems.
In the problems with periodic boundaries (as in the decay test of the
Alfv\'{e}n wave in \S 3.2), a fast Poisson solver based on the fast Fourier
transform can be used.
In the nonlinear simulation of the Parker instability in \S 3.3 with
reflection boundaries along one direction, the computational domain is
doubled to that direction and the resulting boundaries are enforced to be
periodic.
In the two-dimensional shock tube tests in \S 3.1.2, doubling the
computational domain in both directions also makes the resulting
boundaries periodic.
Note that in Eq.~(\ref{FD}) $\phi$'s are coupled with those at every
other cell in a column and row. 
So the (extended) computational domain is divided into four sub-domains,
and $\phi$'s are computed in those sub-domains separately.

\section{TESTS}

In this section we present the results of three tests.  
The first and the second are isothermal versions of 
MHD shock tubes and decay of an Alfv\'{e}n wave, respectively (RJ; RJF).
The shock tube test shows the ability of the IMHD code
to handle all the three MHD wave family structures, while the decay test of an
Alfv\'{e}n wave measures numerical dissipation.
The third test is the simulation of a real 
astrophysical situation, the nonlinear evolution of the Parker
instability under a uniform gravity.
In all the tests, we set the isothermal speed $a=1$.

\subsection{Shock Tube Tests}

Based on the work of RJ, we have devised four shock tube problems
which include discontinuities and rarefaction waves of IMHDs.
To confirm the validity of our numerical solutions we have
compared them to the analytic solutions obtained with an exact, nonlinear
MHD Riemann solver described in RJ.
That Riemann solver iterates from an initial guess of the solution for
the full set of MHD waves based on the given left and right states.
Iteration continues until the solutions
to the innermost wave zone reached from the two opposite 
directions agree within some specified limit (in practice, a relative
error $10^{-5}$). The four shock tube solutions that we have applied in the
test described below are listed in Table 1.
$C_{\rm cour}=0.8$ and $\epsilon_1=0.3$ (for fast modes), $\epsilon_2=0.0$
(for Alfv\'{e}n modes), $\epsilon_3=0.3$ (for slow modes) have been used
in the shock tube test calculations.

\subsubsection{One-Dimensional Shock Tube Tests}

The one-dimensional simulations of the shock tube problems
have been done with 512 cells in a computational tube bounded by $x=[0,1]$. 
We plot in following figures the resulting $\rho$, $B_y$, $B_z$, $v_x$, $v_y$,
and $v_z$ at each cell with open circles and the analytic solutions with lines.

Figure~1a shows the result of the first shock tube problem 
at $t=0.1$ with the initial condition of  
a left state $(\rho=1, v_x=0, v_y=0, v_z=0, B_y=5/\sqrt{4\pi}, B_z=0)$,
a right state $(\rho=0.1, v_x=0, v_y=0, v_z=0, B_y=2/\sqrt{4\pi}, B_z=0)$,
and $B_x=3/\sqrt{4\pi}$.
It exhibits the capturing of a fast rarefaction wave, 
a slow rarefaction wave, a slow shock, and a fast shock whose structures
are plotted in the figure from left to right.
There is no contact discontinuity.
The fast and slow shocks are resolved sharply within several cells.  
In order to see the capturing rotational discontinuities, we have set up the
initial condition of the second shock tube problem as:
a left state $(\rho=1.08, v_x=1.2, v_y=0.01, v_z=0.5, 
                                   B_y=3.6/\sqrt{4\pi},B_z=2/\sqrt{4\pi})$,
a right state $(\rho=1, v_x=0, v_y=0, v_z=0, 
                                   B_y=4/\sqrt{4\pi},B_z=2/\sqrt{4\pi})$,
and $B_x=2/\sqrt{4\pi}$.  Figure~1b shows the result at $t=0.2$.
There are two fast shocks propagating outmost,
and two slow shocks interior to those. Two rotational discontinuities lie
between the fast and slow shocks.
Here the strong fast shocks are resolved sharply, but the weak
slow shocks and rotational discontinuities spread over more cells.
The third shock tube problem has been set up with the initial condition of
a left state $(\rho=0.12, v_x=24, v_y=0, v_z=0, 
                                   B_y=3/\sqrt{4\pi},B_z=0)$,
a right state $(\rho=0.3, v_x=-15, v_y=0, v_z=0, 
                                   B_y=0,B_z=3/\sqrt{4\pi})$,
and $B_x=0$.  Two oppositely moving magnetosonic shocks and a tangential
discontinuity at $t=0.2$ are shown in Figure~1c. 
The magnetosonic shocks are again 
resolved sharply, but the tangential discontinuity spreads over
$\sim 20$ cells.  In the fourth shock tube problem, the initial condition
has been set up with
a left state $(\rho=1, v_x=-1, v_y=0, v_z=0,
                                   B_y=1,B_z=0)$,
a right state $(\rho=1, v_x=1, v_y=0, v_z=0,
                                   B_y=1,B_z=0)$,
and $B_x=0$.
The result at $t=0.16$ in Figure~1d shows two oppositely-moving identical
magnetosonic rarefactions.

\subsubsection{Two-Dimensional Shock Tube Tests}

The two-dimensional simulations of the shock tube problems
have been done with $256 \times 256$ cells in a computational domain bounded 
by $x=[0,1]$ and $y=[0,1]$.  Initially, the domain is divided into two parts
by a diagonal line joining the two points (0,1) and (1,0).  The left state
of the initial conditions for the one-dimensional shock tube problems
has been assigned to the lower left part and the right state to the
upper right part.  The generated structures, including discontinuities 
and rarefactions, propagate parallel to the other diagonal line joining
the two points (0,0) and (1,1).

In Figures~2a and 2b, two-dimensional correspondences of Figures~1a and 1b
are plotted.
In the figures, the following subscripts are used:
$\parallel$ for parallel components of velocity and magnetic field
along the diagonal line joining the two points (0,0) and (1,1),
$\perp$ for perpendicular components but still in the computational plane,
and $z$ for components out of the plane.
Although the resolution of the two-dimensional simulations,
$256 \times 256$ cells, is lower than that of the one-dimensional ones,
512 cells, we see all the structures have been captured correctly.

\subsection{Decay of an Alfv\'{e}n Wave}

RJF carried out a test of the decay of linear waves in order to estimate
numerical dissipations (resistivity and viscosity) in their
adiabatic MHD code.  Following the same idea, the decay of 
a linear Alfv\'{e}n wave has been calculated and numerical dissipation in
our IMHD code has been estimated.
The IMHD equations for viscous and resistive fluid can be written as
\begin{equation}
\frac{\partial \rho}{\partial t} + \nabla \cdot (\rho v) = 0,
\end{equation}
\begin{equation}
\frac{\partial v}{\partial t} + v \cdot \nabla v
+ \frac{1}{\rho} \nabla (a^2 \rho)
- \frac{1}{\rho} (\nabla \times B) \times B
= \frac{1}{\rho} \partial_k \sigma_{ik},
\end{equation}
\begin{equation}
\frac{\partial B}{\partial t}
- \nabla \times (v \times B) = \eta \nabla^2 B.
\end{equation}
In the momentum equation, the viscosity tensor $\sigma_{ik}$ is given by, 
\begin{equation}
\sigma_{ik} =  \mu (  \partial_k v_i + \partial_i v_k 
                   - \case{2}{3} \delta_{ik} \nabla \cdot v)
              + \zeta \delta_{ik} \nabla \cdot v,
\end{equation}
where $\mu$ and $\zeta$ are the dynamic
shear and bulk viscosity, and $\eta$ is the electrical
resistivity.
Under uniform density, $\rho_0$, and uniform magnetic field, 
$B=B_0\hat{x}$, the complex angular frequency of Alfv\'{e}n waves
is predicted from the linear analysis to be
\begin{equation}
\omega = \frac{i}{2}\left(\frac{\mu}{\rho_0}+\eta\right)k^2
\pm c_{\rm A} k 
\left[ 1 - \frac{1}{4c_{\rm A}^2}
\left(\frac{\mu}{\rho_0}-\eta\right)^2 k^2\right]^{1/2},
\end{equation}
where $c_{\rm A}=\sqrt{B_0^2 /2\rho_0}$ is the Alfv\'{e}n 
speed along the wave propagation direction and
$k=(k_x^2+k_y^2)^{1/2}$ is the total wavenumber.
Note that the complex angular frequency of the isothermal Alfv\'{e}n waves is 
exactly the same as that of the adiabatic ones (RJF).
We define the decay rate as
\begin{equation}
\Gamma_{\rm A} = \frac{1}{2}\left(\frac{\mu}{\rho_0} + \eta\right) k^2.
\end{equation}

For the decay test of a linear Alfv\'{e}n wave with the
IMHD code, we have set up an initial condition such that,
$\rho_0=1$, $\delta v_z = v_{\rm{amp}} c_{\rm A} \sin (k_x x + k_y y)$, 
$B = 1 \cdot \hat{x}$, and all other quantities are equal to zero. 
The calculations have been done in a square periodic box with size $L=1$
using from $8 \times 8$ cells to $128 \times 128$ cells
by increasing twice the number of cells in each direction.  
We set $k_x=k_y=2\pi/L$.  Numerical parameters used
are $\epsilon_1=\epsilon_2=\epsilon_3=0$ and $C_{\rm cour}=0.9$ (the
result is not sensitive to these values).  
Figure~3a shows the decay of the Alfv\'{e}n 
wave calculated with $32 \times 32$ cells.  By fitting the peak points
of the decay pattern with respect to time, we have estimated decay rate.
In Figure~3b the resulting normalized decay rates as well as Reynolds
numbers (see RJF for definition) are shown.  Numerical Reynolds numbers
scale almost as $R\propto n_{\rm cell}^2$ indicating the code has a
second-order accuracy.  Compared to the adiabatic MHD code, the IMHD code
has smaller (up to 50\%) numerical dissipation.
This is partly because the IMHD code has one less mode (entropy mode).

\subsection{Parker Instability under a Uniform Gravity}

Nonlinear development of the Parker instability under a point-mass 
dominated gravity was simulated by Matsumoto and his collaborators 
(Matsumoto \etal 1988; Matsumoto \etal 1990; Matsumoto \& 
Shibata 1992). And recently, Basu \etal (1996, 1997) simulated 
the nonlinear evolution of the Parker instability under a uniform gravity. 
As the final test of our IMHD code, we have also followed the nonlinear
evolution of the Parker instability under the uniform gravity.
By comparing our results with those in Basu \etal (1996),
the code's ability to handle a practical problem of astrophysics can be
proved.

Since the Parker system is assumed initially to be in an isothermal 
equilibrium, an IMHD code is a natural tool for simulations.
In the IMHD equations the externally given gravity -$g{\hat z}$ is 
treated as a source term and placed on the right hand side of 
Eq.~(\ref{3dimhd})
with the source vector defined by 
$S=(0,0,0,-gv_z,0,0,0)^T$.  
The gravity has the $z$-component only, so the source 
term is evaluated only when the state vector is updated along the $z$-axis
\begin{equation}
\frac{\partial q}{\partial t} + \frac{\partial F_z}{\partial z} = S.
\end{equation}
Since we use the Strang-type directional splitting in order to reduce
multi-dimensional problems to one-dimensional ones, we also use the same
technique to split the hyperbolic system with a source term into two 
parts,
\begin{eqnarray}
\mbox{Part A:} & \; & \frac{\partial q}{\partial t} +
                     \frac{\partial F_z}{\partial z} = 0,  \\
\mbox{Part B:} & \; & \frac{\partial q}{\partial t} = S.
\end{eqnarray}
Part A is solved by the TVD algorithm, and Part B by a
forward-time-difference. 
To minimize a `splitting error', Part A and Part B are solved by a BAB 
sequence with time step size 0.5$\Delta t$ for Part B and $\Delta t$ for Part
A. The step size $\Delta t$ is determined from the Courant condition.

The Parker system composed of isothermal gas and
magnetic field ${\hat y}B_0(z)$
takes under the uniform gravity an equilibrium configuration given by 
\begin{equation}\label{istate}
\frac{\rho_0(z)}{\rho_0(0)} =
\frac{B_0^2(z)}{B_0^2(0)} =
\exp(-z/H),
\end{equation}
where the gas scale height is $H\equiv(1+\alpha)a^2/g$ and the initial 
ratio of the magnetic to the gas pressure $\alpha(\equiv B_0^2/
[2\rho_0 a^2])$ is assumed a constant.  We have chosen 
$\alpha=1.0$ in the simulation.

The computation domain covers $0 \leq y \leq 12H$ and $0 \leq z \leq 12H$. 
According to the linear stability analysis $12H$ is the horizontal wavelength 
corresponding to the maximum growth rate (Parker 1966). 
Periodic condition has been used in the $y$-boundaries, while reflecting
condition in the $z$-boundaries.
The density scale height $H$, the isothermal sound speed $a$, 
the initial midplane density $\rho_0(0)$, and the initial midplane 
field strength $B_0(0)$ have been chosen as the units of length, velocity, 
density, and magnetic field, respectively.

To initiate the instability we have added random velocity perturbations to the 
equilibrium profile of Eq.~(\ref{istate}). 
Standard deviation of the perturbation 
velocity is $10^{-4}a$ for each of the velocity components. 
To check whether the system follows, in the initial stage, 
the prescription of the linear analysis, 
we plot the logarithmic values of the root-mean-square velocity against time. 
In Figure~4a  
the dotted and dashed lines are for the horizontal and vertical components of 
the velocity, respectively. In the same figure the solid line 
represents the linear growth of a rate 0.34, which is the maximum growth 
rate of the system. At the very early stage the system undergoes a transient 
phase of adjustment, and then quickly develops the Parker instability at the 
rate predicted by the linear analysis. The linear growth gets saturated near 
$t\simeq40$. 

The whole development of the Parker instability may be divided into three 
phases: The linear phase lasts up to $t\simeq$40, from then on the system 
undergoes  
the nonlinear phase until $t\simeq57$, and finally it reaches the damping 
oscillatory phase. Iso-contours and grey maps for density (left
panels) and magnetic field lines and velocity vectors (right panels),
in Figure 4b, present the snap shots of the 
system at the end of the linear phase ($t=40$), at the end of the
nonlinear phase ($t=57$), and finally at $t=80$ of
the damping oscillatory phase.

In the linear phase the perturbations grow predominantly in the upper 
region. In the nonlinear phase the perturbations gradually move towards 
the midplane.
Through the linear and nonlinear phases, our simulation renders features that 
closely agree with those of Basu \etal (1996).
As more matter accumulates, already compressed gas in the valley gets 
over-compressed. The increased gas pressure bounces the valley matter back to 
the upper region, and at the same time, the built-up pressure at the valley 
gets somewhat eased off. This in turn brings the matter back to the valley. 
The system now enters the oscillatory phase of the Parker instability. 
As the field lines are pushed deep down to the valley by the weight of 
over-lying matter, the curvature of the lines becomes small 
to the degree that magnetic field lines undergo reconnection.
Due to the reconnection the matter drops down off 
the reconnected line, thereby the matter is allowed to move across 
the magnetic field line. The field line is now relieved from the burden of 
supporting the gas against the external gravity, and floats upwards. 
On the other hand,
the field line located just below the reconnected one has to support more 
weight than before. Consequently this line now gets reconnected. In this way a 
redistribution of matter with respect to the field lines continues to occur, 
until there is no more reconnection. The system finally settles in an 
equilibrium. Since the reconnection drives the system 
to violate the flux-freezing condition, the final configuration of the system 
is different from that of the Mouschovias equilibrium (1974).  

\section{CONCLUSIONS}

We have developed one- and multi-dimensional numerical codes to solve the
IMHD equations, which are isothermal analogues of the previous 
adiabatic codes (RJ; RJF).  Both the isothermal and adiabatic codes
are based on the same scheme, an explicit finite-difference scheme on 
an Eulerian grid called TVD, which is a second-order-accurate extension of 
the Roe-type upwind scheme.  The shock tube tests
have showed that both codes capture correctly all the structures in MHDs. 
From the decay test of a linear Alfv\'{e}n wave, 
we have found that numerical dissipation of the IMHD code is 
somewhat smaller than that of the adiabatic MHD code.

The robustness of the adiabatic code has been demonstrated through  
the simulations of MHD flows such as the Kelvin-Helmholtz instability 
(Frank \etal 1996; Jones \etal 1997) and jets (Frank \etal 1998;
Jones \etal 1998),
and that of the isothermal code has been done through
the simulation of the Parker instability under the uniform gravity
in this paper.  Furthermore, both codes are fast enough to simulate
multi-dimensional, astrophysical MHD flows using
modest computational resources. Both codes run at about 
400 MFlops on a Cray C90 processor (RJF), and the isothermal code
updates zones about twice as fast as the adiabatic code.      
Together with the adiabatic code, the isothermal code is a useful
tool to study the nonlinear evolution of astrophysical MHD flows.

\acknowledgments

JK was supported by the Ministry of Science and Technology
through Korea Astronomy Observatory grant 97-5400-000.  
The work by DR was supported in part by KOSEF through the 1997
Korea-US Cooperative Science Program 975-0200-006-2.
The work by TWJ was supported in part by the NSF through grants
AST93-18959, INT95-11654 and AST96-16964, by NASA grant NAG5-5055
and by the University of Minnesota Supercomputing Institute.
SSH was supported in part by the Ministry of 
Education, Basic Science Research Institute grant No. BSRI-97-5411.

\clearpage

\centerline{\bf FIGURES}

\figurenum{1a}
\figcaption{
One-dimensional IMHD shock tube test.
The initial condition is
$(\rho, v_x, v_y, v_z, B_y,           B_z)=$
$(1,    0,   0,   0,   5/\sqrt{4\pi}, 0  )$
in the left region,
$(\rho, v_x, v_y, v_z, B_y,           B_z)=$
$(0.1,  0,   0,   0,   2/\sqrt{4\pi}, 0  )$ 
in the right region,
$B_x=3/\sqrt{4\pi}$ and $a=1$ for the whole computational interval.
Open circles represent the numerical solution, while
lines represent the analytic solution with an exact nonlinear Riemann solver.
The calculation has been done with 512 cells.
A snapshot at $t=0.1$ shows from left to right 
(1) fast rarefaction, (2) slow rarefaction,
(3) slow shock, and (4) fast shock. }

\figurenum{1b}
\figcaption{
One-dimensional IMHD shock tube test.
The initial condition is 
$(\rho, v_x, v_y,  v_z, B_y, B_z)=$
$(1.08, 1.2, 0.01, 0.5, 3.6/\sqrt{4\pi}, 2/\sqrt{4\pi})$
in the left region,
$(\rho, v_x, v_y,  v_z, B_y,             B_z)=$
$(1,    0,   0,    0,   4/\sqrt{4\pi},   2/\sqrt{4\pi})$
in the right region,
$B_x=2/\sqrt{4\pi}$ and $a=1$ for the whole computational interval. 
Open circles represent the numerical solution, while
lines represent the analytic solution with an exact nonlinear Riemann solver.
The calculation has been done with 512 cells.
A snapshot at $t=0.2$ shows from left to right 
(1) fast shock, (2) rotational discontinuity, (3) slow shock, 
(4) slow shock, (5) rotational discontinuity, and (6) fast shock. }

\figurenum{1c}
\figcaption{
One-dimensional IMHD shock tube test.
The initial condition is
$(\rho, v_x, v_y, v_z, B_y,           B_z)=$
$(0.12, 24,  0,   0,   3/\sqrt{4\pi}, 0  )$
in the left region,
$(\rho, v_x, v_y, v_z, B_y,           B_z)=$
$(0.3,  -15, 0,   0,   0,             3/\sqrt{4\pi})$ 
in the right region,
$B_x=0$ and $a=1$ for the whole computational interval. 
Open circles represent the numerical solution, while
lines represent the analytic solution with an exact nonlinear Riemann solver.
The calculation has been done with 512 cells.
A snapshot at $t=0.2$ shows from left to right 
(1) magnetosonic shock, (2) tangential discontinuity,
and (3) magnetosonic shock. }

\figurenum{1d}
\figcaption{
One-dimensional IMHD shock tube test.
The initial condition is
$(\rho, v_x, v_y, v_z, B_y, B_z)=$
$(1,    -1,  0,   0,   1,   0  )$
in the left region,
$(\rho, v_x, v_y, v_z, B_y, B_z)=$
$(1,    1,   0,   0,   1,   0  )$ 
in the right region,
$B_x=0$ and $a=1$ for the whole computational interval. 
Open circles represent the numerical solution, while
lines represent the analytic solution with an exact nonlinear Riemann solver.
The calculation has been done with 512 cells.
A snapshot at $t=0.16$ shows from left to right 
(1) magnetosonic rarefaction, and (2) magnetosonic rarefaction. }

\figurenum{2a}
\figcaption{
Two-dimensional IMHD shock tube test.
The initial condition is
$(\rho, v_{\parallel}, v_{\perp}, v_z, B_{\perp},     B_z)=$
$(1,   0,             0,         0,   5/\sqrt{4\pi}, 0  )$
in the lower left region,
$(\rho, v_{\parallel}, v_{\perp}, v_z, B_{\perp},     B_z)=$
$(0.1,  0,             0,         0,   2/\sqrt{4\pi}, 0)$ 
in the upper right region,
$B_{\parallel}=3/\sqrt{4\pi}$ and $a=1$ in the whole computational domain.  
Open circles represent the numerical solution, while
lines represent the analytic solution with an exact nonlinear Riemann solver.
The calculation has been done with $256\times256$ cells.
The structures shown at $t=0.1 \sqrt{2}$ along a diagonal line joining 
the two points (0,0) to (1,1) are same as those of Figure~1a.
}

\figurenum{2b}
\figcaption{
Two-dimensional IMHD shock tube test.
The initial condition is
$(\rho, v_{\parallel}, v_{\perp}, v_z, B_{\perp},       B_z)=$
$(1.08, 1.2,           0.01,      0.5, 3.6/\sqrt{4\pi}, 2/\sqrt{4\pi})$
in the lower left region,
$(\rho, v_{\parallel}, v_{\perp}, v_z, B_{\perp},     B_z)=$
$(1,    0,             0,         0,   4/\sqrt{4\pi}, 2/\sqrt{4\pi})$ 
in the upper right region,
$B_{\parallel}=2/\sqrt{4\pi}$ and $a=1$ in the whole computational domain.  
Open circles represent the numerical solution, while
lines represent the analytic solution with an exact nonlinear Riemann solver.
The calculation has been done with $256\times256$ cells.
The structures shown at $t=0.2 \sqrt{2}$ along a diagonal line joining 
the two points (0,0) to (1,1) are same as those of Figure~1b.
}

\figurenum{3a}
\figcaption{
Time evolution of $<\delta B_z^2>^{1/2}$ and $<\delta v_z^2>^{1/2}$ 
in the decay test of a two-dimensional Alfv\'{e}n wave.  Initially, a
standing Alfv\'{e}n wave has been set up in a computational domain
with $32 \times 32$ cells, and its oscillation has been followed.
}

\figurenum{3b}
\figcaption{
Normalized decay rate, $\Gamma_{\rm A} L/c_A$, and magnetic Reynolds number,
$R$, as a function of the number of cells along one direction of the
computation domain.
At a given resolution, the peak-to-peak decay rate of the root-mean-square
of $z$-velocity (top) and the corresponding Reynolds
number (bottom) are plotted with filled circles, respectively.    
The calculations have been done with 
$8 \times 8$, $16 \times 16$, $32 \times 32$,   
$64 \times 64$, and $128 \times 128$ cells.   
For comparison, dotted lines of $(\Gamma_{\rm A} L/c_{\rm A}) 
\propto n_{\rm cell}^{-2}$
and $R \propto n_{\rm cell}^{2}$ are drawn.
}

\figurenum{4a}
\figcaption{
Time evolution of the root-mean-square of the horizontal velocity,
$<v_y^2>^{1/2}$, and the vertical velocity, $<v_z^2>^{1/2}$,
in a simulation of the Parker instability under a uniform gravity.
The magnetohydrostatic equilibrium state together
with random velocity perturbations has been given as an initial condition of
the simulation in the computational domain of $256 \times 256$ cells. 
The solid line represents the predicted maximum linear 
growth with perturbation wavelength $\lambda_y = 12$, and 
$\lambda_z/2=12$.
The normalization units are the isothermal sound speed and the scale height.  
}

\figurenum{4b}
\figcaption{
Evolution of the Parker instability under a uniform gravity.
At three time epochs $t=40$ (top), $t=57$ (middle), and $t=80$ (bottom), 
grey maps of density together with equi-density lines are plotted in left 
panels, and the velocity vectors with magnetic field lines in right panels. 
The values of the ten equi-density lines are the initial 
exponential densities at $z=1, \cdots, 10$. 
Magnetic field lines are chosen so that the magnetic flux 
between two consecutive lines is constant.
At each time epoch, the unit of the velocity vectors is represented.
}

\clearpage

\begin{deluxetable}{cccccc}
\footnotesize
\tablenum{1a}
\tablecaption{Shock Tube Test 1a \label{tab1a}}
\tablehead{ \colhead{$\rho$} & \colhead{$v_x$} & \colhead{$v_y$} &
\colhead{$v_z$} & \colhead{$B_y$} & \colhead{$B_z$} }
\startdata
 1.0000E+00& 0.0000E+00& 0.0000E+00& 0.0000E+00& 1.4105E+00& 0.0000E+00\nl
 5.7648E-01& 9.3200E-01&-5.3737E-01& 0.0000E+00& 5.9825E-01& 0.0000E+00\nl
 3.0968E-01& 1.3718E+00&-1.0767E-02& 0.0000E+00& 7.8902E-01& 0.0000E+00\nl
 1.2358E-01& 7.2565E-01&-7.6338E-01& 0.0000E+00& 9.0720E-01& 0.0000E+00\nl
 1.0000E-01& 0.0000E+00& 0.0000E+00& 0.0000E+00& 5.6419E-01& 0.0000E+00\nl
\enddata
\tablenotetext{}{with $B_x = 8.4628E-01$}
\end{deluxetable}

\begin{deluxetable}{cccccc}
\footnotesize
\tablenum{1b}
\tablecaption{Shock Tube Test 1b \label{tab1b}}
\tablehead{ \colhead{$\rho$} & \colhead{$v_x$} & \colhead{$v_y$} &
\colhead{$v_z$} & \colhead{$B_y$} & \colhead{$B_z$} }
\startdata
 1.0800E+00& 1.2000E+00& 1.0000E-02& 5.0000E-01& 1.0155E+00& 5.6419E-01\nl
 1.5087E+00& 6.4673E-01& 1.3132E-01& 5.6740E-01& 1.4677E+00& 8.1542E-01\nl
 1.5087E+00& 6.4673E-01& 2.4196E-01& 3.0857E-01& 1.6036E+00& 4.9750E-01\nl
 1.7451E+00& 6.0765E-01& 7.3388E-02& 2.5628E-01& 1.4736E+00& 4.5716E-01\nl
 1.3560E+00& 5.4030E-01&-2.1440E-01& 1.6699E-01& 1.6825E+00& 5.2198E-01\nl
 1.3560E+00& 5.4030E-01&-1.2262E-01&-6.1311E-02& 1.5757E+00& 7.8783E-01\nl
 1.0000E+00& 0.0000E+00& 0.0000E+00& 0.0000E+00& 1.1284E+00& 5.6419E-01\nl
\enddata
\tablenotetext{}{with $B_x = 5.6419E-01$}
\end{deluxetable}

\clearpage

\begin{deluxetable}{cccccc}
\footnotesize
\tablenum{1c}
\tablecaption{Shock Tube Test 1c \label{tab1c}}
\tablehead{ \colhead{$\rho$} & \colhead{$v_x$} & \colhead{$v_y$} &
\colhead{$v_z$} & \colhead{$B_y$} & \colhead{$B_z$} }
\startdata
 1.2000E-01& 2.4000E+01& 2.3130E-16& 0.0000E+00& 8.4628E-01& 4.4409E-16\nl
 1.7079E+00& 9.2149E-02& 2.3130E-16& 0.0000E+00& 1.2045E+01& 6.3206E-15\nl
 4.1960E+00& 9.2149E-02& 0.0000E+00& 0.0000E+00& 7.2478E-16& 1.1837E+01\nl
 3.0000E-01&-1.5000E+01& 0.0000E+00& 0.0000E+00& 0.0000E+00& 8.4628E-01\nl
\enddata
\tablenotetext{}{with $B_x = 0.0000E+00$}
\end{deluxetable}

\begin{deluxetable}{cccccc}
\footnotesize
\tablenum{1d}
\tablecaption{Shock Tube Test 1d \label{tab1d}}
\tablehead{ \colhead{$\rho$} & \colhead{$v_x$} & \colhead{$v_y$} &
\colhead{$v_z$} & \colhead{$B_y$} & \colhead{$B_z$} }
\startdata
 1.0000E+00&-1.0000E+00& 0.0000E+00& 0.0000E+00& 1.0000E+00& 0.0000E+00\nl
 4.6392E-01& 7.8159E-08& 0.0000E+00& 0.0000E+00& 4.6392E-01& 0.0000E+00\nl
 1.0000E+00& 1.0000E+00& 0.0000E+00& 0.0000E+00& 1.0000E+00& 0.0000E+00\nl
\enddata
\tablenotetext{}{with $B_x = 0.0000E+00$}
\end{deluxetable}

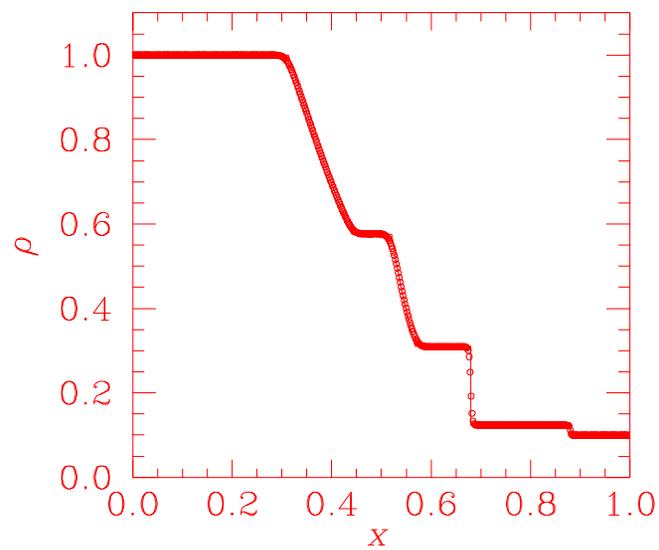
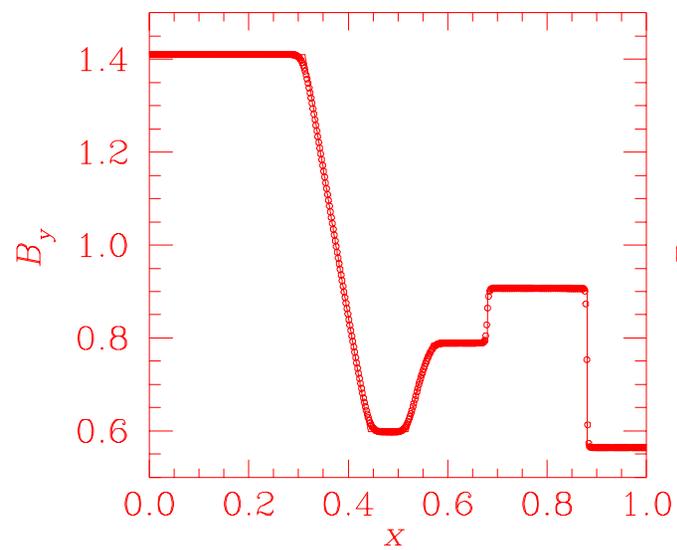
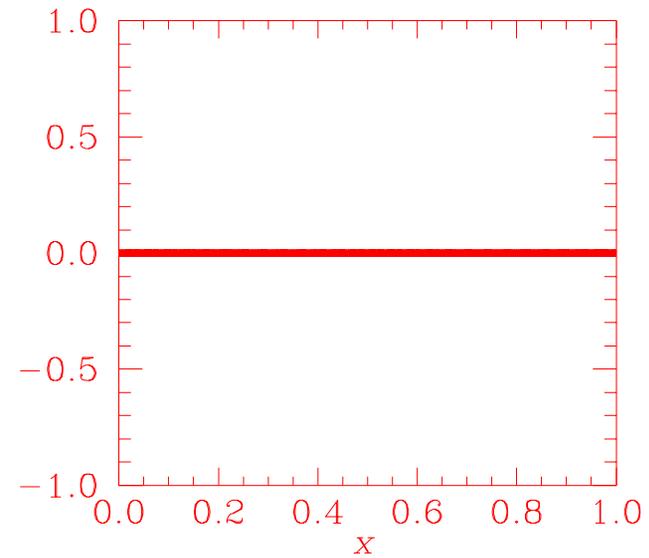
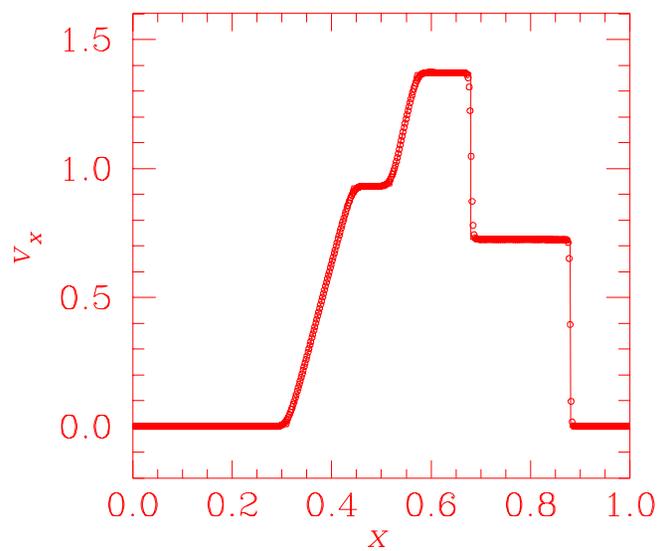
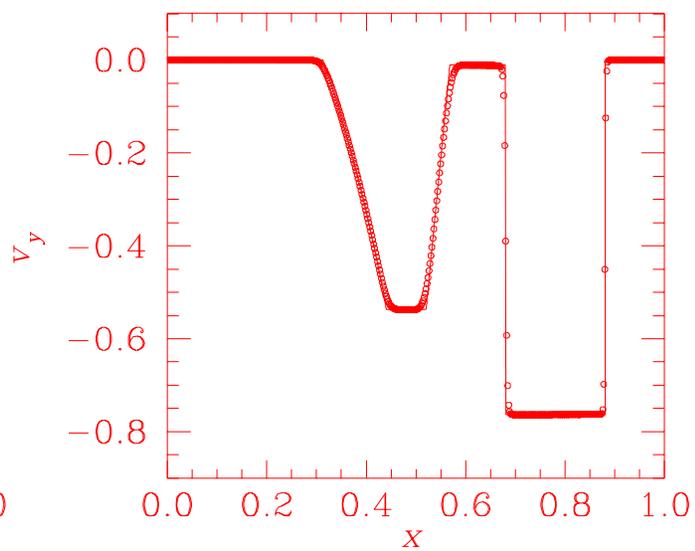
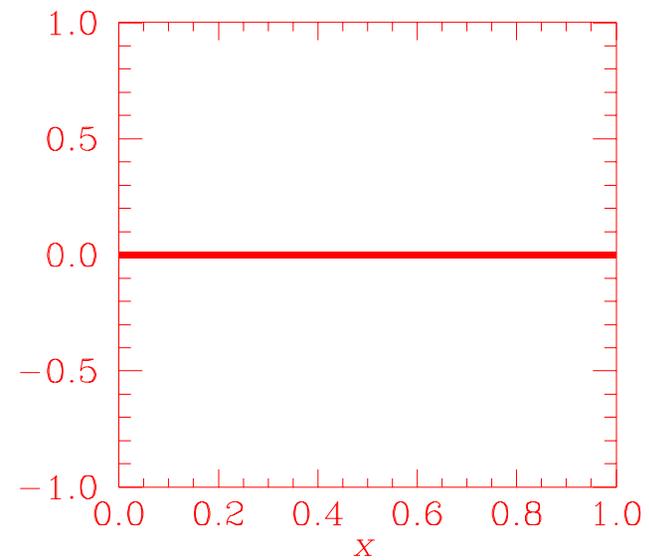
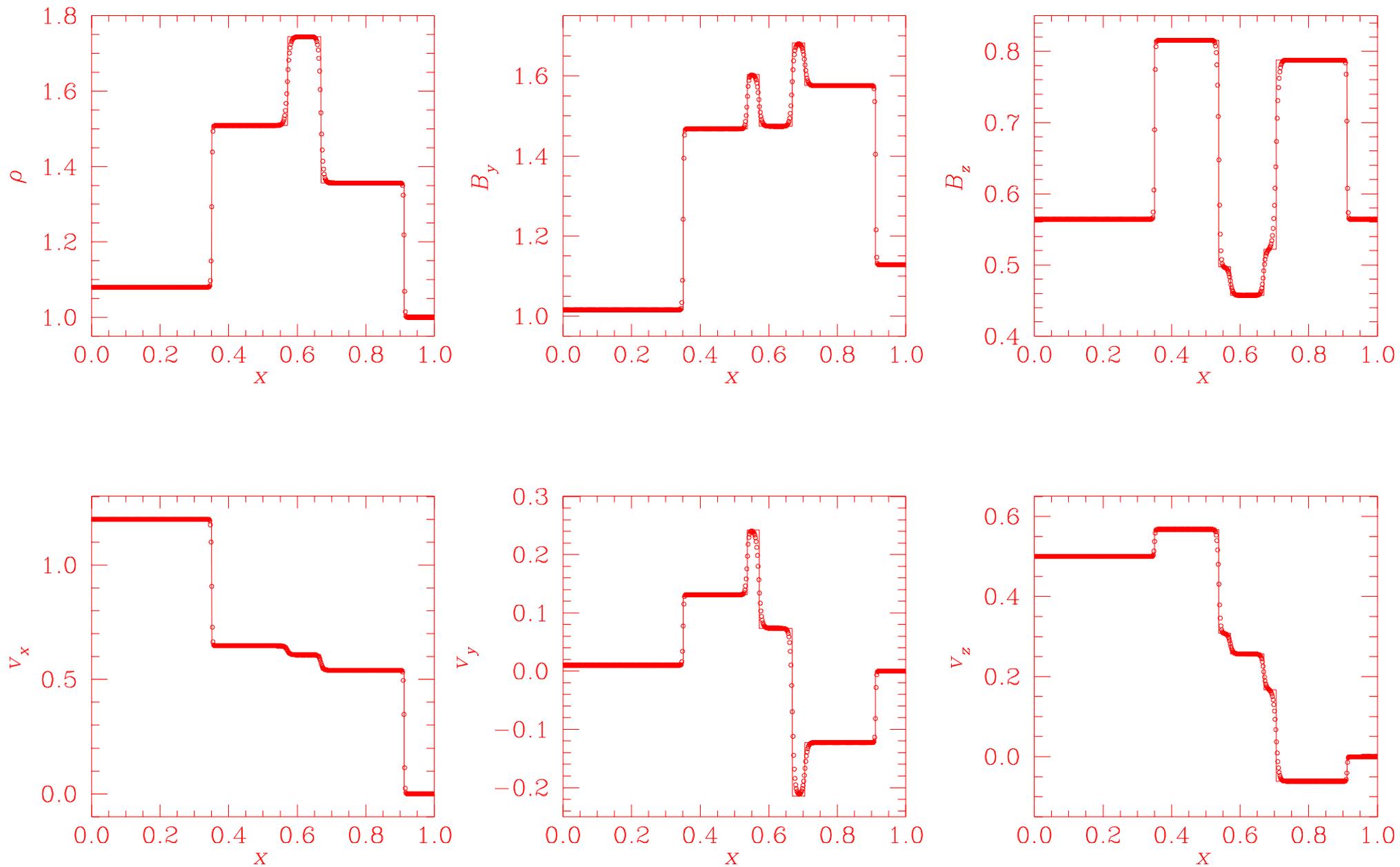
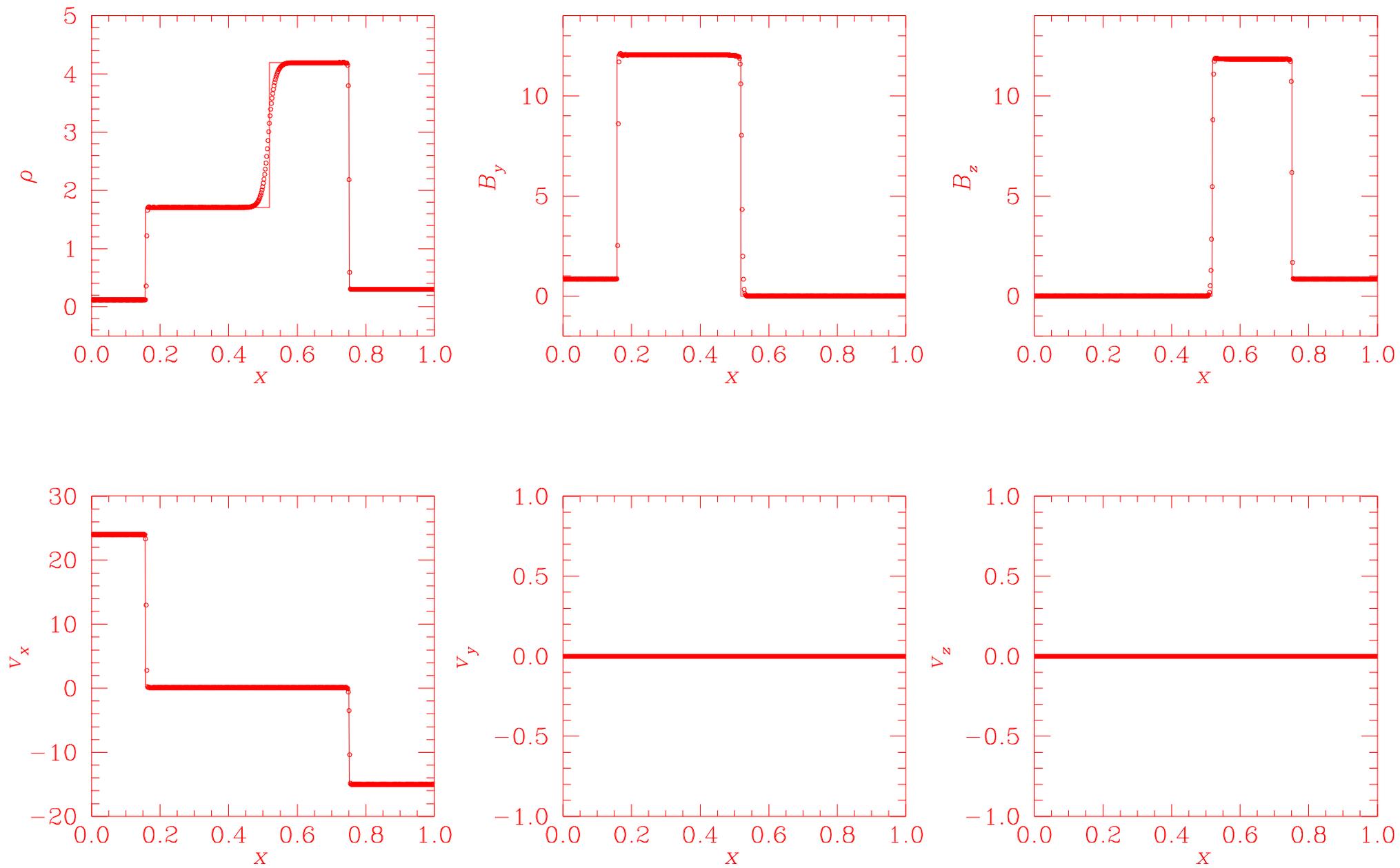
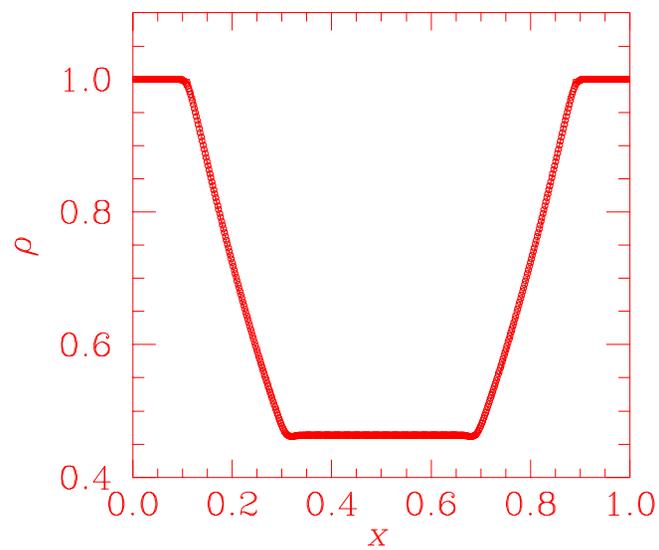
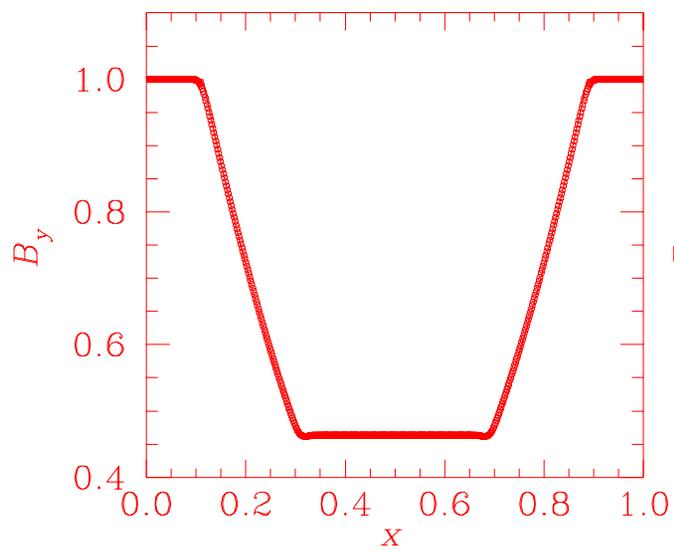
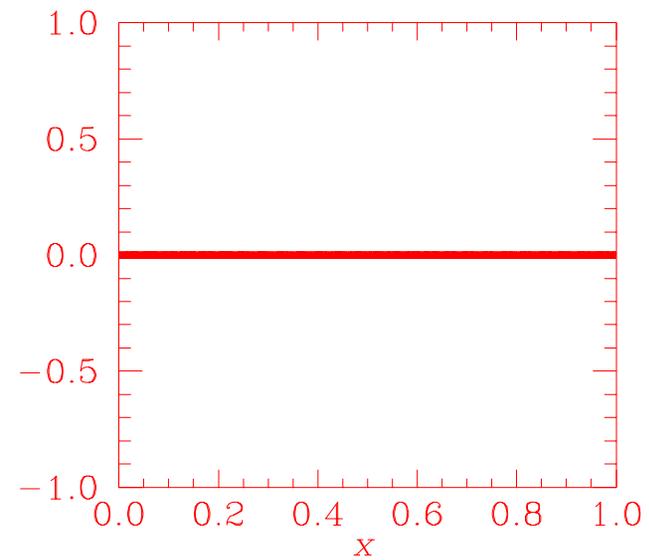
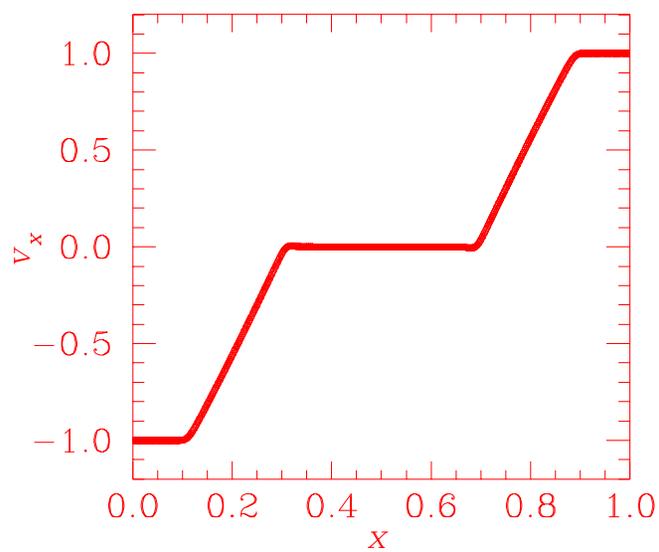
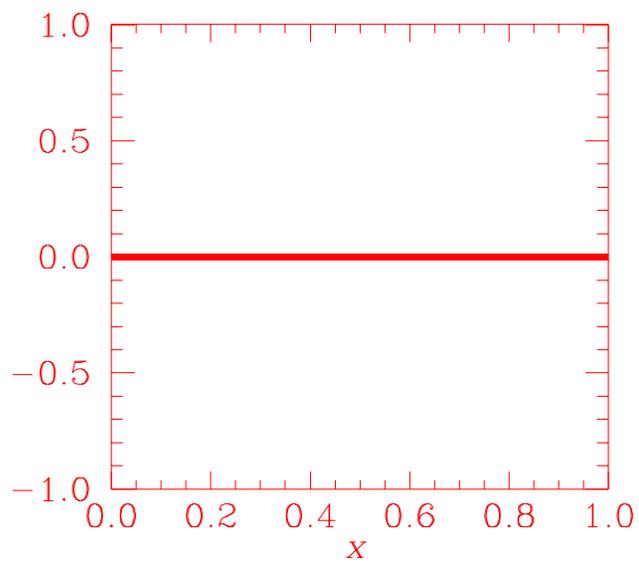
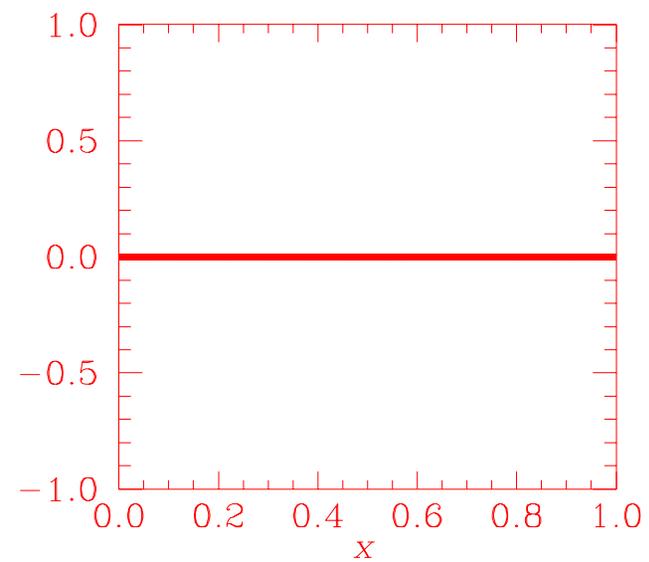
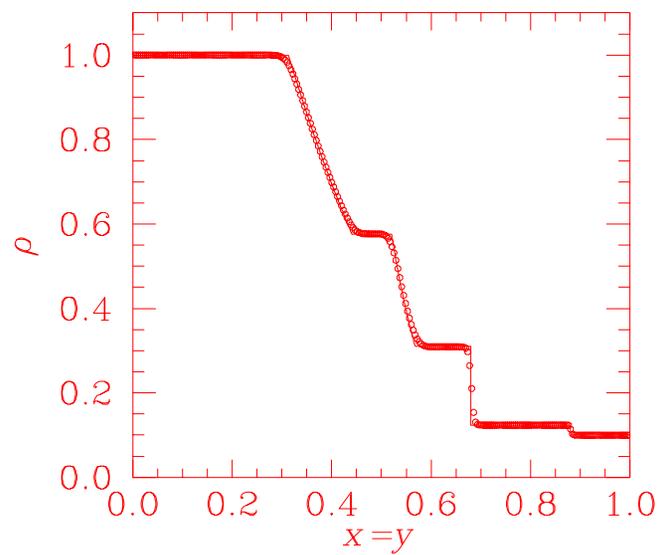
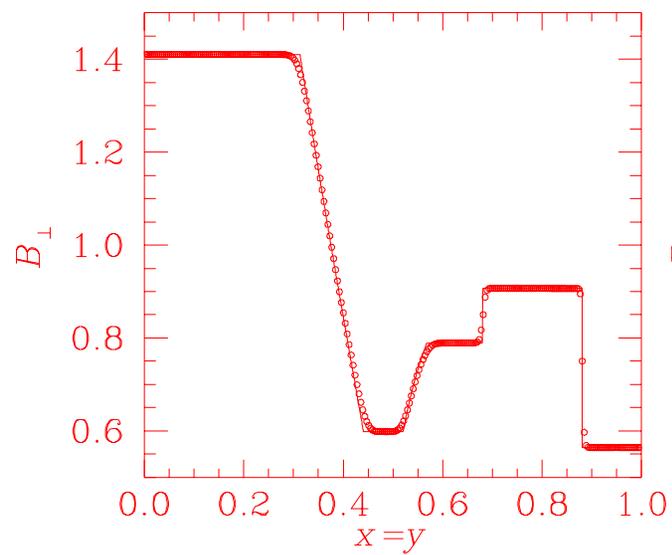
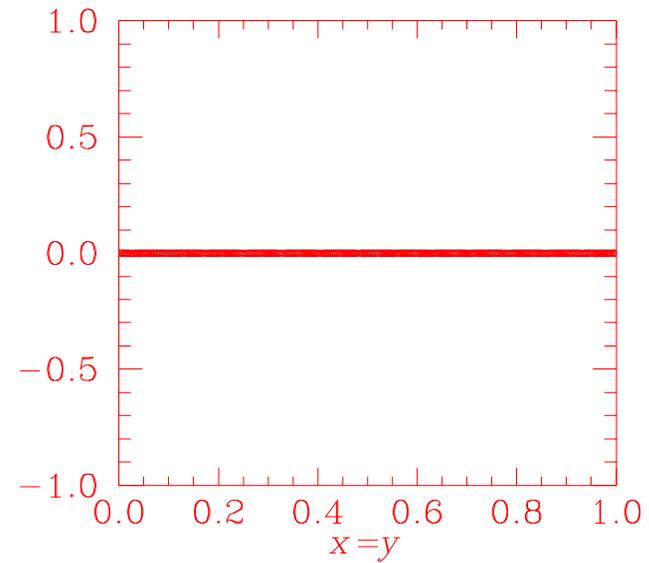
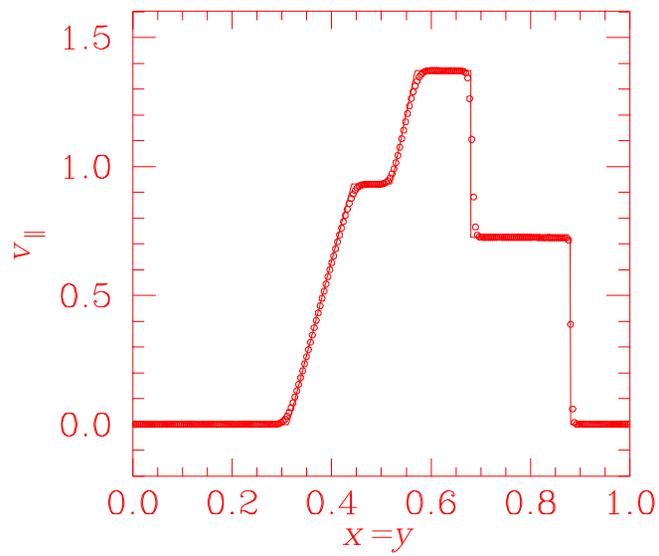
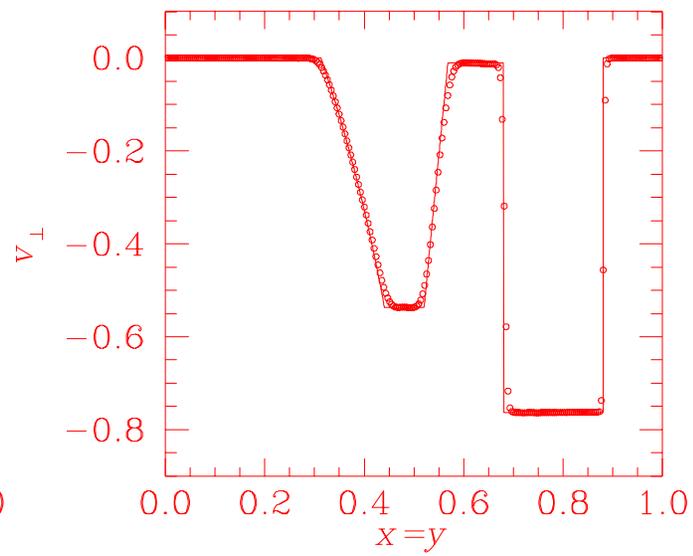
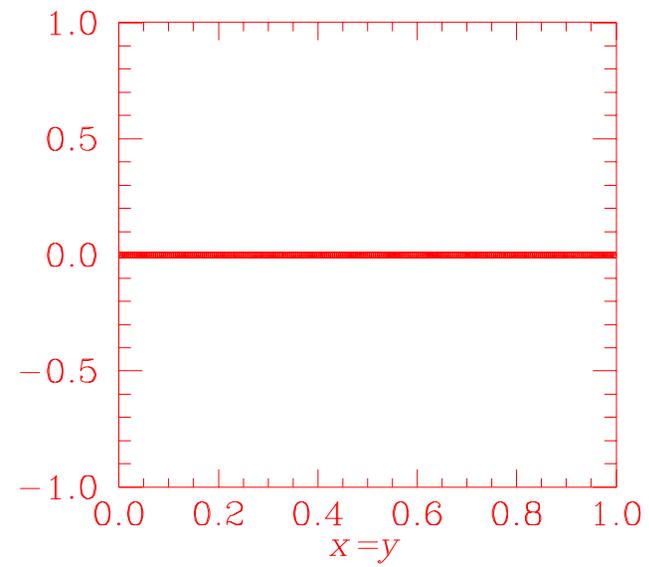
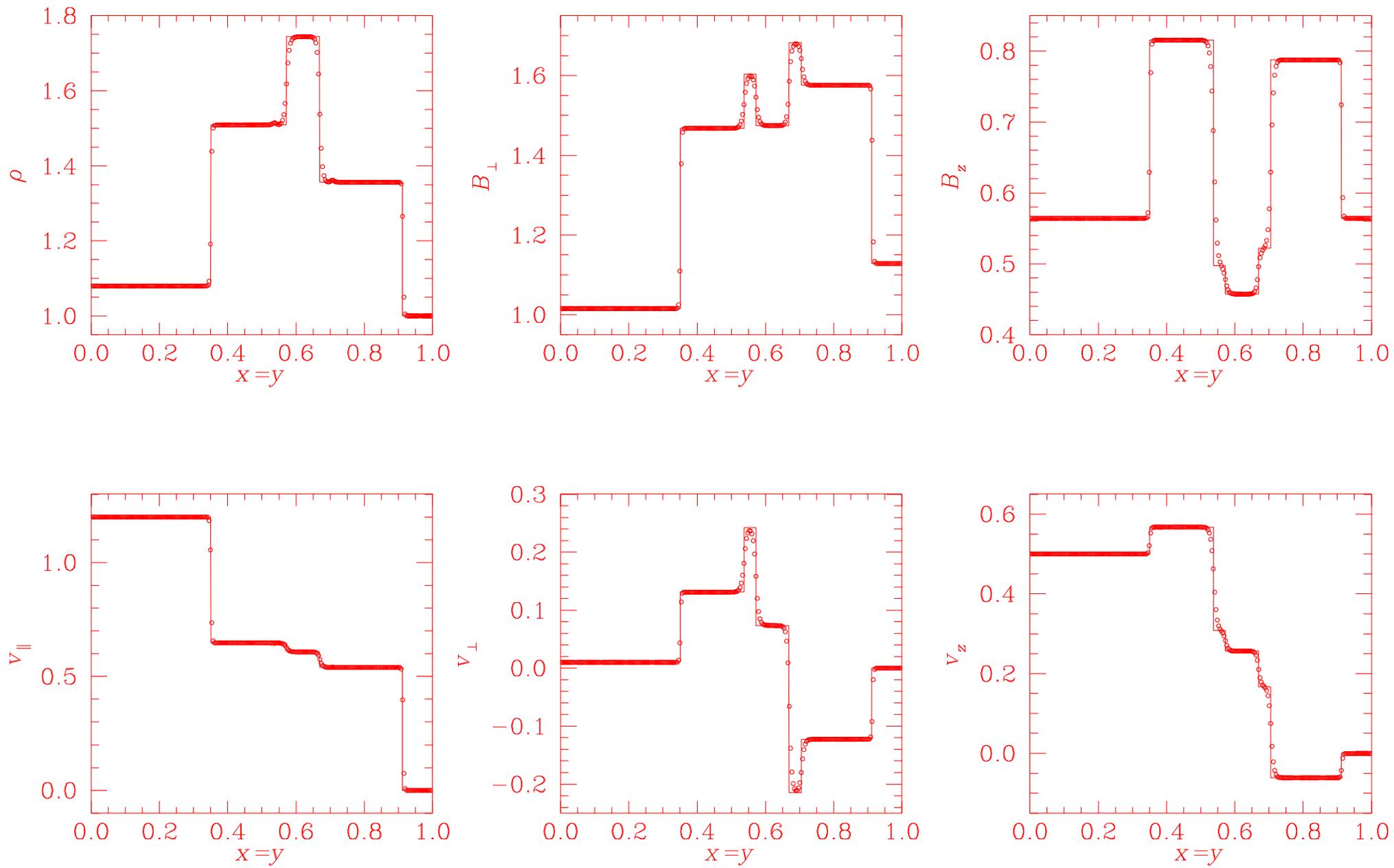
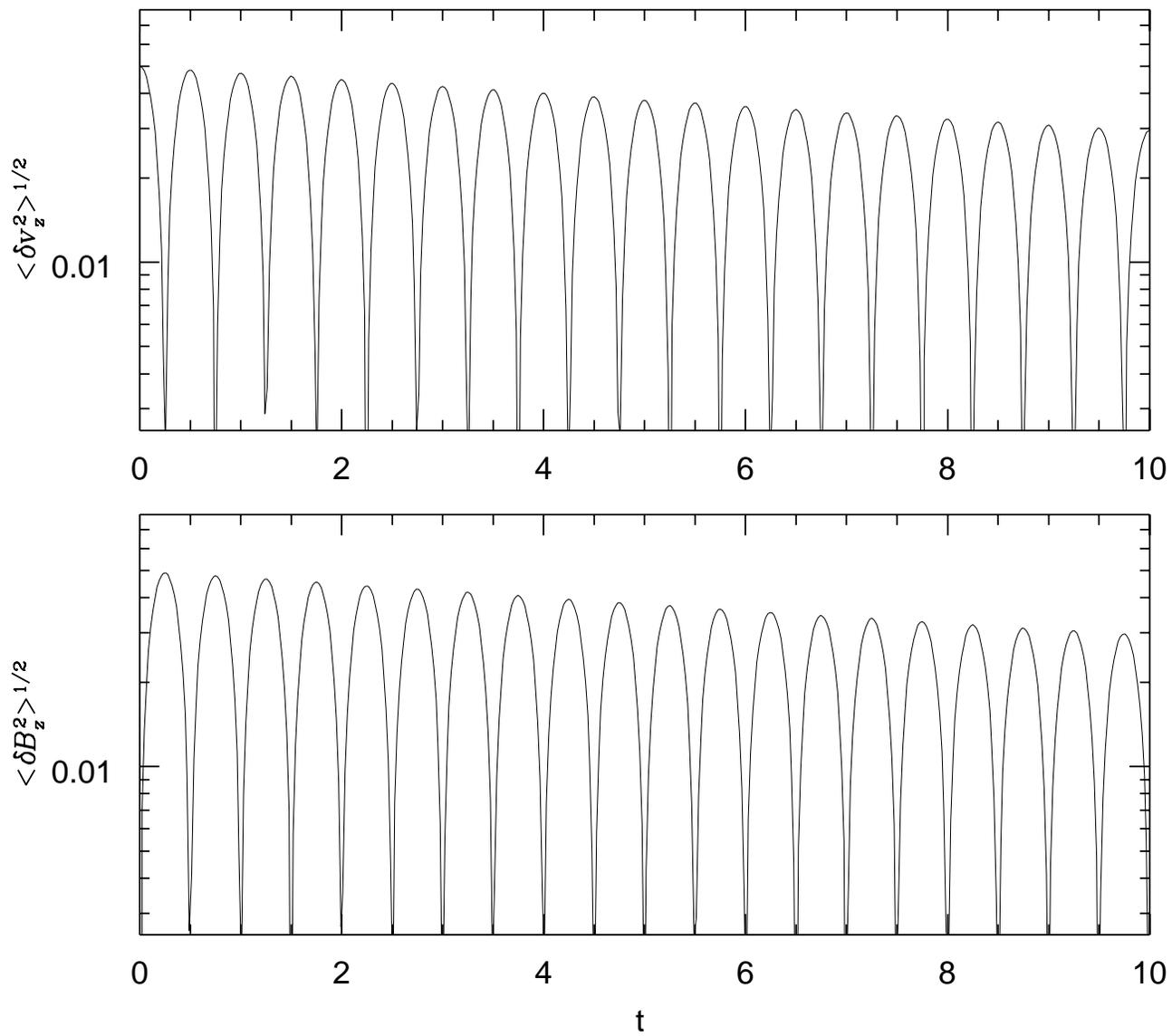
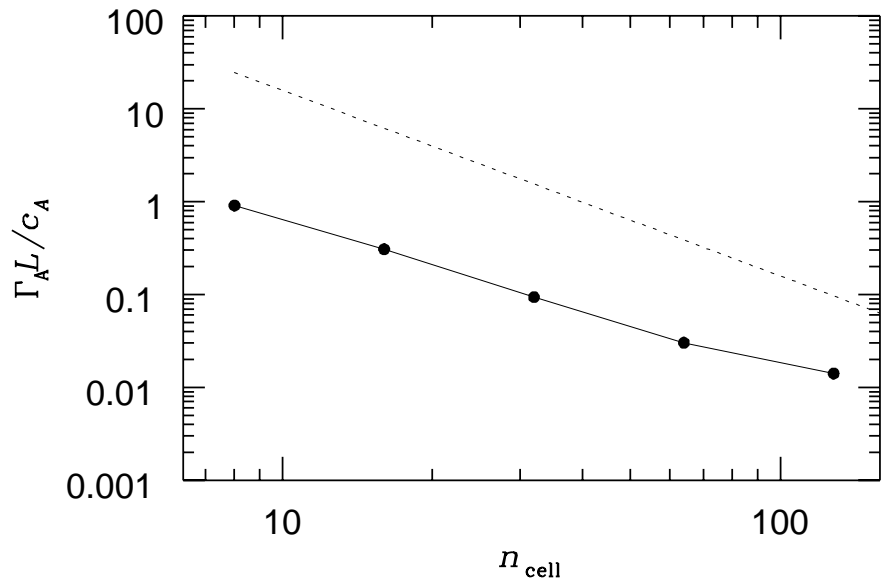
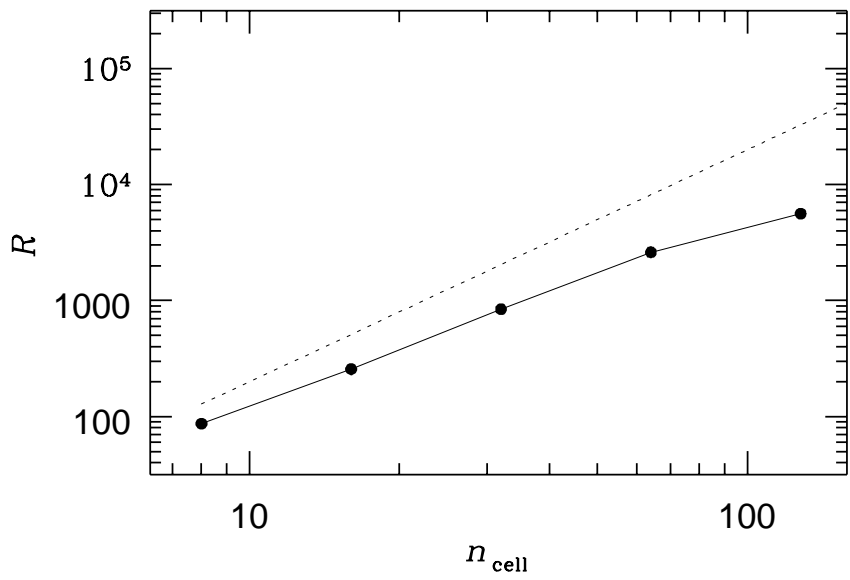
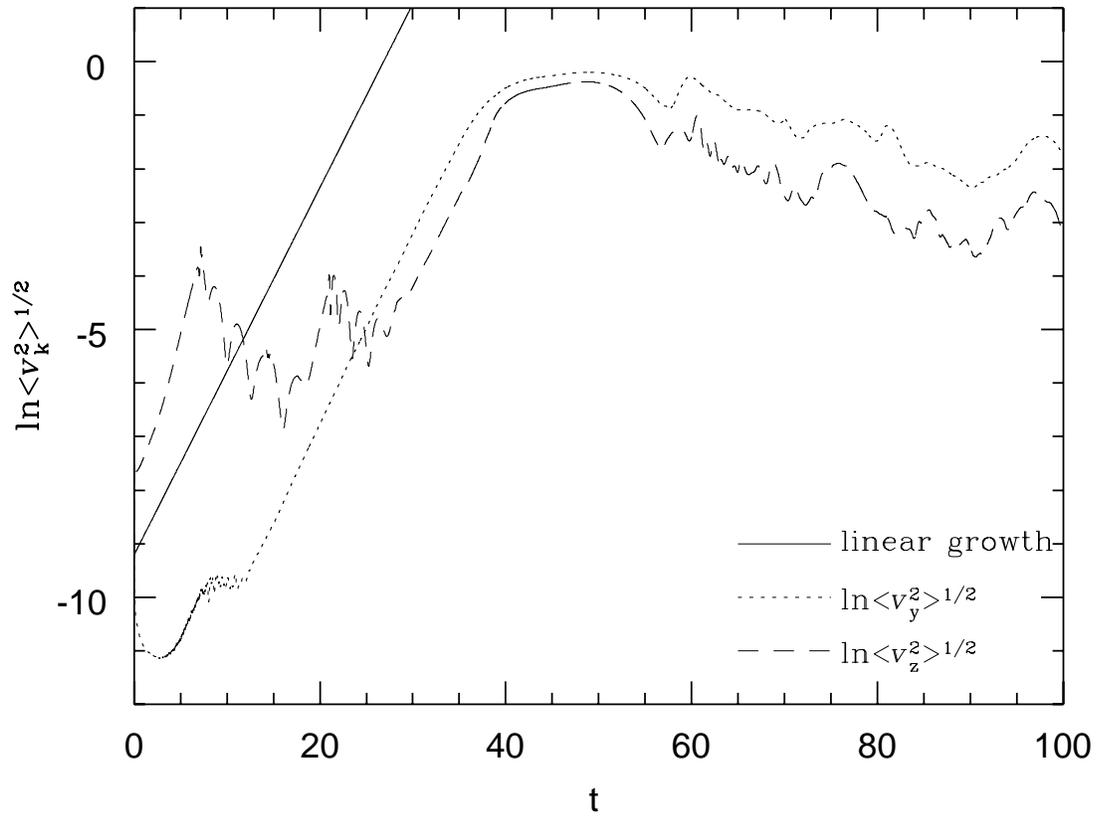
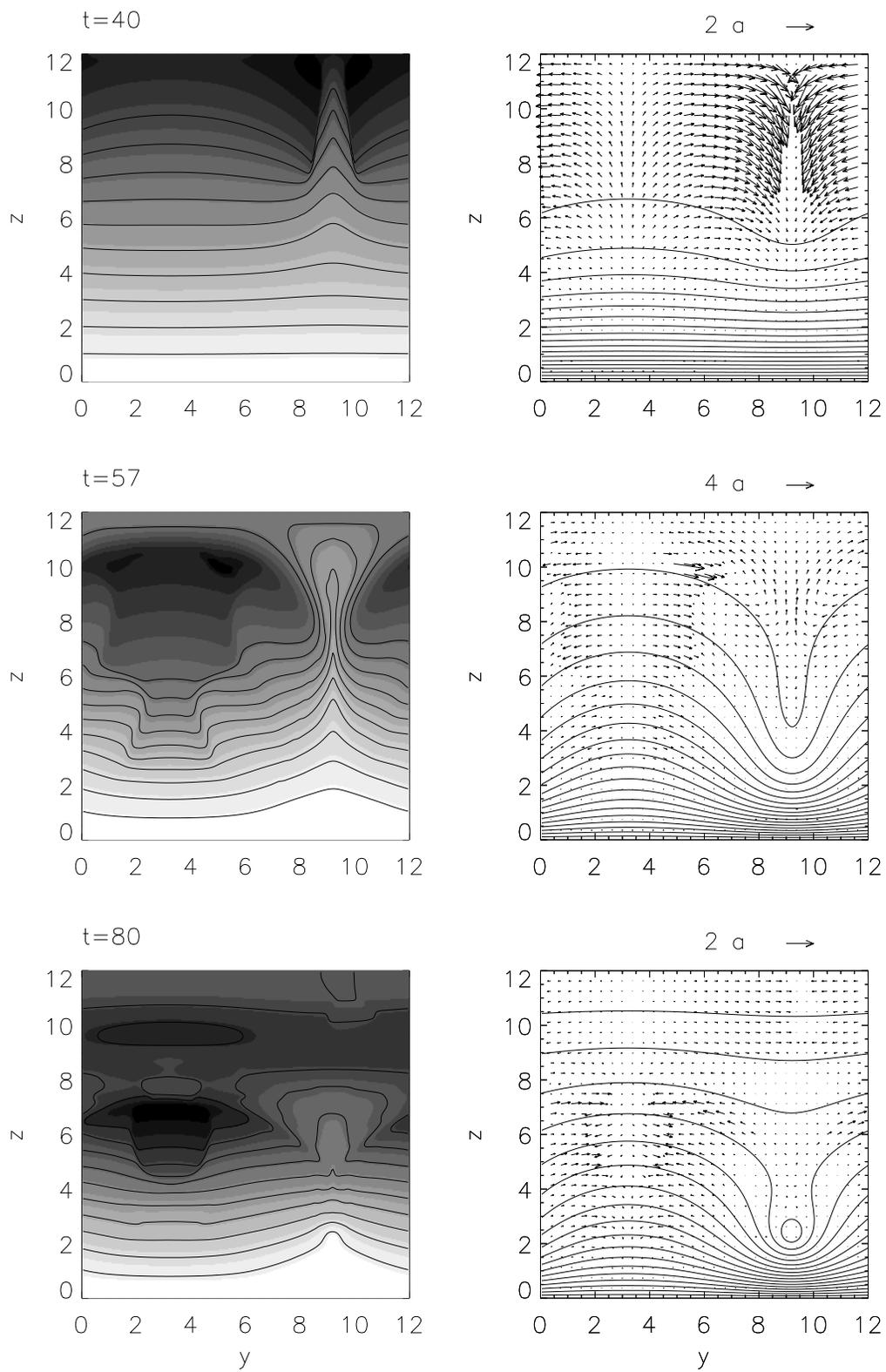
\end{document}